\documentclass[mzfinal]{mzpaper}

 \usepackage{amsmath}			
 \usepackage{amsfonts}			
 \usepackage{array}             
 \usepackage{setspace}
 \usepackage{threeparttable}
 \usepackage{booktabs}
 \usepackage{graphicx}			
 \usepackage{subfigure}         
 \usepackage{caption2}
 \usepackage{color}			    
\usepackage{hyperref}

 \usepackage{units}
 
\begin{document}

\volume{1} \volnumber{1} \pubmonth{January} \pubyear{2000}
\doctype{1}{}
\spage{001} \epage{001} \doisuffix{0000}
\price{0.00}

\title[Benchmarking in a rotating annulus]{Benchmarking in a rotating annulus: a comparative experimental and numerical study of baroclinic wave dynamics}
\author[M. Vincze et al.]{
\noindent Miklos Vincze$^1$,
Sebastian Borchert$^2$, 
Ulrich Achatz$^2$, 
Thomas von Larcher$^3$,
Martin Baumann$^4$,
Claudia Hertel$^5$,
Sebastian Remmler$^6$,
Teresa Beck$^4$,
Kiril Alexandrov$^1$,
Christoph Egbers$^1$,
Jochen Fr\"ohlich$^5$,
Vincent Heuveline$^4$,
Stefan Hickel$^6$,
\textnormal{and} Uwe Harlander$^1$\corrauthor{uwe.harlander@tu-cottbus.de}
}

\affiliation{
$^1$Lehrstuhl f\"ur Aerodynamik und Str\"omungslehre, Brandenburgische Technische Universit\"at Cottbus-Senftenberg, Cottbus, Germany\\
$^2$Institut f\"ur Atmosph\"are und Umwelt, Goethe-Universit\"at at Frankfurt am Main, Frankfurt am Main, Germany\\
$^3$Institut f\"ur Mathematik, Freie Universit\"at Berlin, Berlin, Germany\\
$^4$Engineering Mathematics and Computing Lab (EMCL), Interdisziplin\"ares Zentrum f\"ur Wissentschaftliches Rechnen, Universit\"at Heidelberg, Heidelberg, Germany\\
$^5$Institut f\"ur Str\"omungsmechanik, Technische Universit\"at Dresden, Dresden, Germany\\
$^6$Lehrstuhl f\"ur Aerodynamik und Str\"omungsmechanik, Technische Universit\"at M\"unchen, M\"unchen, Germany\\
}

\maketitle

\begin{abstract}
The differentially heated rotating annulus is a widely studied tabletop-size laboratory model
of the general mid-latitude atmospheric circulation. The two most relevant factors of cyclogenesis, namely rotation
and meridional temperature gradient are quite well captured in this simple arrangement. The radial temperature difference in the cylindrical tank and its rotation rate can be set so that the isothermal surfaces in the bulk tilt, leading to the formation of baroclinic waves. The signatures of these waves at the free water surface have been analyzed via
infrared thermography in a wide range of rotation rates (keeping the radial temperature difference constant) and
under different initial conditions. 
In parallel to the laboratory experiments, five groups of the MetStr\"om collaboration
have conducted numerical simulations in the same parameter regime using different approaches and solvers,
and applying different initial conditions and perturbations. The experimentally and numerically obtained baroclinic wave
patterns have been evaluated and compared in terms of their dominant wave modes, spatio-temporal variance properties and drift rates. Thus certain ``benchmarks'' have been created that can later be used as test cases
for atmospheric numerical model validation.
\end{abstract}

\selectlanguage{english}



\section{Introduction}
In the endeavor to improve weather forecasting and climate prediction techniques, the validation and fine-tuning of numerical models of large-scale atmospheric processes play clearly crucial roles. However, in such a complex system as the real atmosphere, validation tests are especially difficult to perform. Besides the issues that arise due to coarse-graining -- a central problem of the numerical modeling of any hydrodynamic problem -- in the case of atmospheric processes the unavoidable imperfection of the governing equations themselves is also a considerable source of inaccuracies. In the commonly applied hydro-thermodynamic equations the unresolved (or even physically not properly understood) processes are either neglected or taken into account via empirical parametrization. Thus, the separation of discretization errors from the ones originating from the theoretical formulation of a given model poses a real challenge to researchers.

Yet, there is a way to carry out systematic and reproducible tests under controlled circumstances, and
to capture a large segment of the complexity of these large-scale flows through relatively simple, tabletop-size
experiments, based on the principle of hydrodynamic similarity. Under laboratory conditions it is possible
to adjust the governing physical parameters and thus to separate different processes that cannot be studied
independently in the real atmosphere. Therefore, laboratory experiments provide a
remarkable test bed to validate numerical techniques and models aiming to investigate geophysical flows.
This was one of the primary goals of the German Science Foundation's (DFG) priority program MetStr\"om. Research focuses on the theory and methodology of multiscale meteorological-fluid mechanics modelling and accompanying reference experiments supported model validation.

One of these reference experiments was the differentially heated rotating annulus. This classical apparatus to study the basic dynamics of the mid-latitude atmosphere has been introduced by \cite{fultz} based on the principles first suggested by \cite{vettin}. The two most relevant factors of cyclogenesis, namely the planetary rotation and the meridional temperature gradient are quite well captured in this simple arrangement.
The set-up (Fig.1)
consists of a cylindrical gap mounted on a turntable and rotating around its vertical axis of symmetry. The
inner side wall of the annulus is cooled whereas the outer one is heated, thus the working fluid experiences a radial temperature gradient. At high enough rotation rates the isothermal surfaces tilt, leading to baroclinic instability. The extra potential energy stored in this unstable configuration is then converted into kinetic energy, exciting drifting wave patterns of temperature and
momentum anomalies. 
The basic underlying physics of such baroclinic waves has been subject of extensive theoretical \citep{eady, mason, lorenz}, numerical \citep{gp_williams,miller_butler,thomas_slope} and experimental \citep{read,christoph_book,thomas_npg,uwe_obst} research throughout the past decades.
Furthermore, some studies focused on the quantitative comparison of temperature statistics \citep{gyure} and propagation dynamics of passive tracers \citep{viki} obtained from annulus experiments and from actual atmospheric data. Even meteorological data assimilation techniques \citep{ravela,young_read_12} and techniques operational in meteorological ensemble prediction \citep{uwe_ait,hoff,young_read_08} have also been studied by using annulus data.

The experimental part of the present study was conducted in the fluid dynamics laboratory of the Brandenburg Technical University at Cottbus-Senftenberg (BTU CS). The infrared thermographic snapshots of the drifting baroclinic waves at the free water surface have been analyzed in a wide range of rotation rates (keeping the radial temperature difference constant) and under different initial conditions.
In parallel to the experiments, five numerical groups of the MetStr\"om collaboration (Goethe University Frankfurt, University of Heidelberg, FU Berlin, TU Dresden and TU Munich)   
have conducted simulations in the same parameter regime using different numerical approaches, solvers and subgrid parametrizations, and applying different initial conditions and perturbations for stability analysis. 
The obtained baroclinic wave
patterns have been evaluated through determining and comparing their statistical variance properties, drift
rates and dominant wave modes. Thus certain ``benchmarks'' are created that can be used as test cases
for atmospheric numerical model validation in the future.
 
Similar comparative studies of experiments and numerical simulations in baroclinic annuli are far not unprecedented: the first such investigations stretch back to the 1980s \citep{james,hignett,read_03,rlh,anthony1,anthony2}, where the comparisons were mostly based on pointwise sub-surface temperature time series. The very same experimental apparatus that was used in the present work set-up has already been used to test and validate subgrid-scale parametrization methods of two of the numerical models also used here (see the paper of \cite{borchert} in the present issue). In another recent comparative study, the effect of the addition of a sloping bottom topography to this set-up was analyzed both experimentally and numerically \citep{our_npg}. However, to the best of our knowledge, the present study is the very first to systematically compare different numerical schemes and two series of experiments with different initial conditions. 

Our paper is organized as follows. Section 2 outlines the experimental set-up, and the experimental and numerical methods used. The results are presented in Section 3. In Section 4 we summarize the results and discuss their implications on the physics of the underlying dynamics.   
    
\section{Methods}
\subsection{Experimental apparatus and procedures}
\begin{figure}
\noindent\includegraphics[width=5cm]{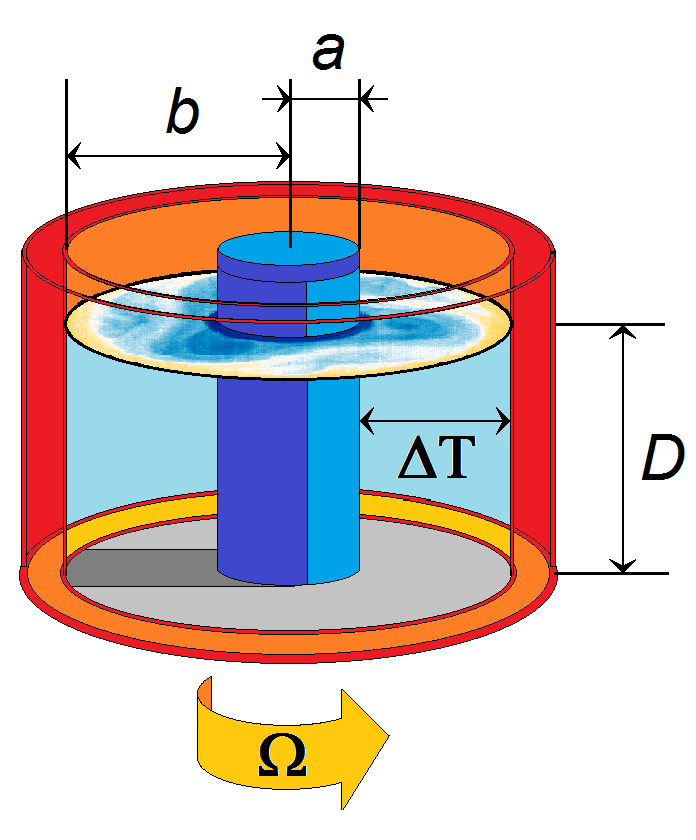}
 \centering
\caption{Schematic drawing of the laboratory set-up. For the values of the geometric parameters shown, see the text. The counter-clockwise direction of rotation is indicated.}
\label{setup_schem}
\end{figure}

The laboratory experiments of the present study have been conducted in the baroclinic wave tank of BTU CS.
This tank was mounted on a turntable, and was divided by coaxial cylindrical sidewalls (Fig. \ref{setup_schem}) into three sections. The innermost compartment (made of anodized aluminum) housed coolant pipes in which cold water was circulated. The temperature in this middle cylinder was monitored via a digital thermometer and kept constant by a thermostat with a precision of $0.05$ K. 
The outermost annular compartment contained heating wires and water as heat conductive medium. Here four thermometers (identical to that of the middle cylinder) provided temperature data for a computer-controlled feedback loop to maintain constant temperature (for the technical details on the applied control methods we refer to the paper of \cite{thomas_npg}. The temperatures in the inner and outer section were set to the values of $18.5\pm 0.25^\circ$C and $26.5\pm 0.25^\circ$C, respectively, yielding a radial temperature difference of $\Delta T=8\pm 0.5$ K. 

The working fluid -- de-ionized water -- occupied the annular gap ranging from $a=4.5$ cm to $b=12$ cm in the radial direction. The water depth was set to $D=13.5$ cm in all experimental runs, thus the vertical aspect ratio of the cavity was $\Gamma=D/(b-a)=1.8$. The water surface was \emph{free} to enable the observation of surface temperature patterns via infrared thermography (the observed wavelength band is generally absorbed by glass or acrylic, thus covering the tank with a rigid lid was not possible). The physical properties of the fluid are characterized by its kinematic viscosity $\nu=1.004\times 10^{-6}$ m$^2$/s and its thermal conductivity $\kappa=0.1434\times 10^{-6}$ m$^2$/s, yielding a Prandtl number of $Pr\equiv\nu/\kappa\approx 7.0$.

\begin{figure}
\noindent\includegraphics[width=8cm]{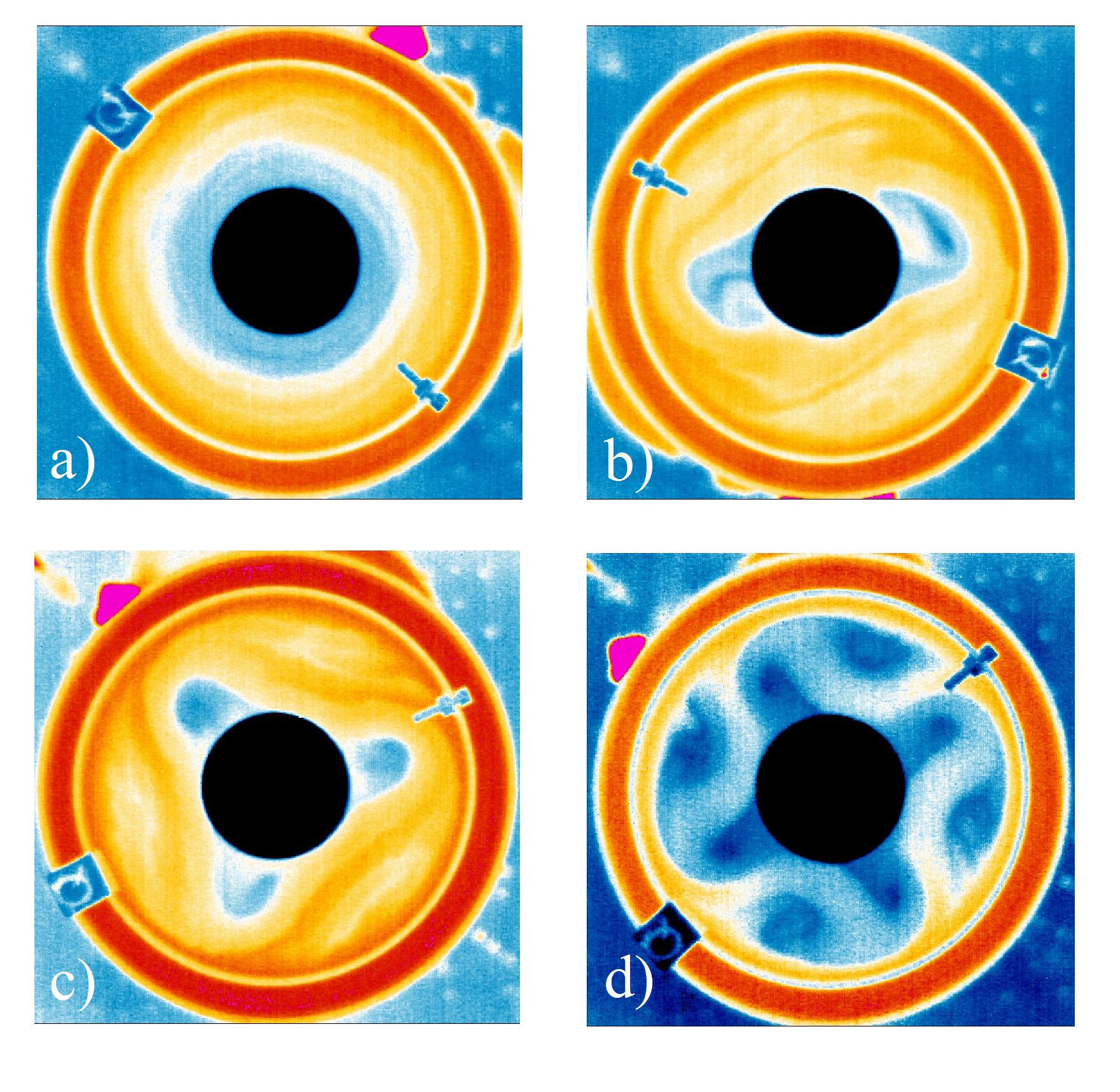}
 \centering
\caption{Four typical thermographic snapshots of surface temperature patterns in the rotating annulus. a) An axially symmetric ($m=0$) pattern  at $\Omega=2.28$ rpm; b) A two-fold symmetric ($m=2$) baroclinic wave at $\Omega=3.23$ rpm; c) $m=3$ at $\Omega=4.20$ rpm; d) $m=4$ at $\Omega=6.16$ rpm.}
\label{exemplary_patterns}
\end{figure}

Since the temperature difference $\Delta T$ as well as the aforementioned geometric and material quantities were kept constant throughout the experiments, rotation rate (i.e. angular velocity) $\Omega$ was the single control parameter to be adjusted between the subsequent runs. The minimum rotation rate investigated was $\Omega_{\rm min}=2.26$ rpm (revolutions per minute), where the flow was found to be axially symmetric, i.e. its radial and vertical structure was independent from azimuthal angle $\theta$, indicating the absence of baroclinic instability. The highest investigated rotation rate was $\Omega_{\rm max}=20.91$ rpm. Here, four-fold symmetric baroclinic wave patterns were observed (see the exemplary thermographic snapshots of Fig. \ref{exemplary_patterns}). Within the interval ranging from $\Omega_{\rm min}$ to $\Omega_{\rm max}$, our measurements were taken at 17 different rotation rates. For each of these cases, two types of initial conditions were applied: the so-called ``spin-up'' and ``spin-down'' sequences. In the former (latter) initialization procedure the target rotation rate $\Omega$ was approached starting from the previously studied smaller (higher) rotation rate $\Omega_i$, with $|\Omega- \Omega_i|\approx 1$ rpm. The rotation rate was then gradually increased (decreased) by $\delta\Omega\approx0.1$ rpm in every 2 minutes; thus, it took approximately $20$ minutes to reach the required $\Omega$ from $\Omega_i$. 10 minutes after arriving to $\Omega$ the data acquisition started and lasted for $40-80$ minutes in each case. Afterwards this gradual increasing (decreasing) procedure of the rotation rate continued to reach the next $\Omega$, with the previous parameter point as $\Omega_i$, providing a long ``spin-up'' (``spin-down'') experiment series. Thus, in total $17\times 2$ measurements were performed and evaluated. Note, that in order to enable the standard initialization procedures at the end parameter points $\Omega_{\rm min}$ and $\Omega_{\rm max}$, the initial value $\Omega_i$ was set smaller than  $\Omega_{\rm min}$ or larger than $\Omega_{\rm max}$, when required. However, no data acquisition took place at these ``out-of-range'' parameter points.       

The infrared camera was mounted above the middle of the tank and was fixed in the laboratory frame (not co-rotating). In every $\Delta t=2$ s, $640\times 480$-pixel thermographic snapshots were taken, providing a precision of around $0.03$ K for temperature differences. The obtained temperature fields can be considered surface temperature patterns, since the penetration depth of the applied wavelength into water is measured in millimeters. The captured snapshots were acquired and stored by a computer, where they were converted to ASCII arrays (by organizing the temperature values from all pixels into matrix format) for further evaluation.   

\begin{figure}
\noindent\includegraphics[width=8cm]{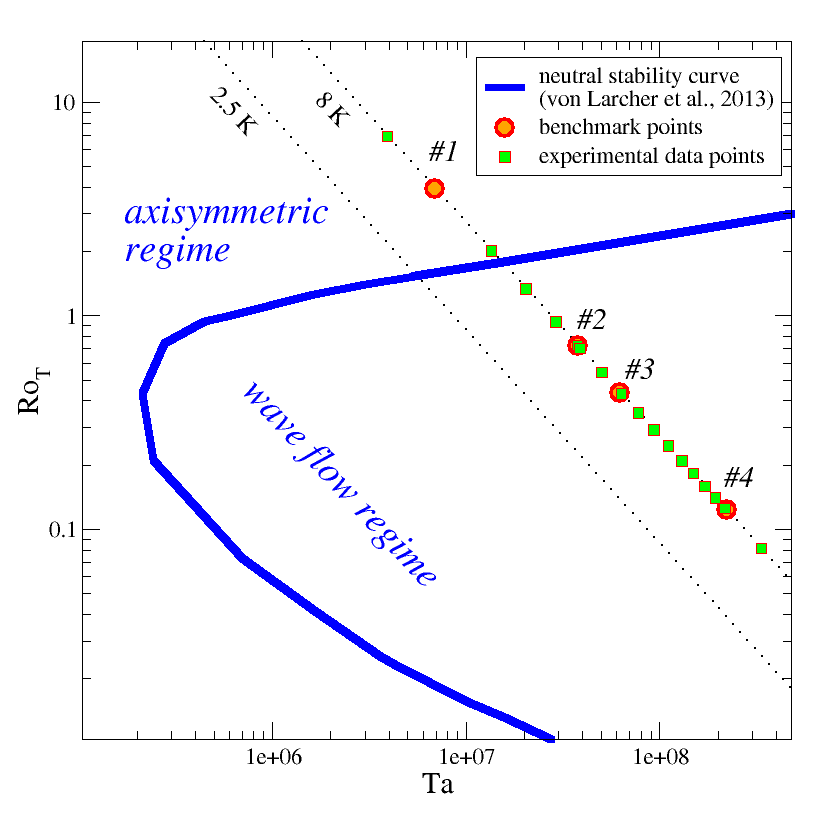}
 \centering
\caption{The neutral stability curve (thick blue line) in the parameter plane of $Ta$ and $Ro_T$, as obtained via linear stability analysis by \cite{thomas_slope}, using the geometrical and material parameters of the BTU C-S wave tank. The line corresponding to the studied radial temperature difference $\Delta T=8$ K is also indicated, along with the experimental data points (squares) and the benchmark parameter points (circles) of the present comparative study, to be discussed later.}
\label{Ta_Ro}
\end{figure}

The most important classic non-dimensional parameters widely used to compare the results obtained from different baroclinic annulus set-ups are the Taylor number $Ta$ and thermal Rossby number $Ro_T$ (also known as Hide number). The former is basically a non-dimensional measure of rotation rate $\Omega$ and reads as
\begin{equation}
Ta=\frac{4\Omega^2(b-a)^5}{\nu^2D},
\label{Ta}
\end{equation}
whereas $Ro_T$ expresses the ratio of the characteristic velocity of the thermally driven flow to the rotation rate in the form of
\begin{equation}
Ro_T=\frac{Dg\alpha\Delta T}{\Omega^2(b-a)^2},
\label{Ro}
\end{equation}
where $g$ is the acceleration due to gravity and $\alpha=2.07\times 10^{-4}$ K$^{-1}$ represents the volumetric thermal expansion coefficient of the working fluid. Note, that $Ta$ and $Ro_T$ are clearly not independent parameters: in the case of the present study where the experiments were conducted at a practically constant value of $\Delta T$, an inverse proportionality $Ro\propto Ta^{-1}$ holds; thus either one of these parameters per se sufficiently describes the applied conditions. Nevertheless, to demonstrate the broader context of the studied domain, we present a conceptual $Ta-Ro_T$ regime diagram in Fig. \ref{Ta_Ro}. The anvil-shaped thick (blue) curve represents the layout of the so-called neutral stability curve (as obtained numerically by \cite{thomas_slope}, to the left of which the flow is axially symmetric (radial ``sideways convection''). To the right of the curve, the emergence of baroclinic wave patterns (as the ones in Fig. \ref{exemplary_patterns}b,c and d) characterizes the flow, which -- for even higher values of $Ta$ -- become irregular in shape as the system approaches geostrophic turbulence (a state not studied in the present paper).         
The curve corresponding to the constant radial temperature difference $\Delta T = 8$ K that lied in the focus of the present work is also indicated (by a dotted curve), along with the experimental parameter points and the four benchmark points (to be addressed later).   
\subsection{Numerical methods}
In this subsection we briefly describe the different numerical models and methods used for the numerical simulations. 
\subsubsection{Governing equations}
The applied numerical models computed approximate solutions of the hydrodynamic equations of motion in the Boussiesq approximation \citep{realvallis}, using different initialization procedures, grids, time steps, boundary conditions and sub-grid--scale parametrization schemes. The overall geometric parameters of the simulation domain corresponded to the aforementioned dimensions of the annular cavity of the laboratory set-up.
The governing equations themselves read as:
\begin{eqnarray}
\frac{\partial\vec{u}}{\partial t}+(\vec{u}\cdot\nabla)\vec{u}&=&-2\Omega\vec{e_z}\times\vec{u}-\frac{1}{\rho_{0}}\nabla p+\\ \frac{\delta\rho}{\rho_0} g\vec{e_z}+\nu\nabla^2\vec{u},\label{eq:momentum}\\
\nabla\cdot\vec{u} & = & 0,\label{equ:continuity}\\
\frac{\partial T}{\partial t}+(\vec{u}\cdot\nabla)T & = & \kappa\nabla^2 T,\label{equ:temperature}
\end{eqnarray}
where $\vec{e_z}$ denotes the unit vector in vertical direction (directed upwards), $\vec{u}$ represents the velocity field, $p$ is the pressure and $\delta\rho$ denotes the difference between the density of the given fluid parcel and the reference density $\rho_0$ (in the Boussinesq approximation $| \delta\rho |\ll\rho_0$ holds). The first term on the right hand side of (\ref{eq:momentum}) accounts for the Coriolis force which -- being an inertial force -- appears in the co-rotating reference frame. This form of the equation was thus used in the implementations of the cylFloit, EULAG, INCA and LESOCC2 models, occasionally also including the terms for centrifugal and Euler forces in (\ref{eq:momentum}), which are generally negligible in the investigated parameter range. For HiFlow$^3$, however, the governing equations were solved in the ``laboratory frame'', hence there the Coriolis term (or any other inertial force term) was virtually absent and the rotation of the tank entered the dynamics through the boundary conditions.

\subsubsection{cylFloit}
The implementation of the \emph{cylindrical flow solver with implicit turbulence model} (cylFloit) is described  
in \cite{borchert}. 
The governing equations are \eqref{eq:momentum} to \eqref{equ:temperature} with slight modifications -- the centrifugal acceleration $\Omega^2 r$  is added to the right-hand side of the radial component of \eqref{eq:momentum} -- in cylindrical coordinates. 
The temperature dependence of the density deviation $\delta\rho(T)$, kinematic viscosity $\nu(T)$ 
and thermal diffusivity $\kappa(T)$ was approximated in the form of second-order polynomial fits to 
empirical reference data for the studied temperature range. Because of this temperature dependence, 
$\nu$ and $\kappa$ depend implicitly on space and time, which is the reason why the viscous stress and the heat 
conduction have slightly different forms than the rightmost terms in \eqref{eq:momentum} and \eqref{equ:temperature}. 
In order to simulate the spin-up and spin-down of the annulus, the Euler acceleration $-(\mathrm{d}\Omega/\mathrm{d} t) r$ 
is added to the right-hand side of the azimuthal component of \eqref{eq:momentum}.
 
The boundary conditions for the temperature were isothermal at the inner and outer sidewalls of the cavity 
(i.e. at radii $r=a$ and $r=b$): 
$T |_{r=a}=T_a\equiv 24^{\circ}$C and $T |_{r=b}=T_b\equiv 32^{\circ}$C, respectively, yielding $\Delta T= 8.0$ K, in agreement with the laboratory set-up. On the top ($z=D$) and bottom ($z=0$) boundaries no-flux conditions were applied for the temperature (i.e. $\nabla T \vec{e_z} | _{z=0,D}\equiv0$). For the velocities at the bottom and lateral sidewalls, no-slip conditions were prescribed ($\vec{u}|_{z=0}=\vec{u}|_{r=a}=\vec{u}|_{r=b}\equiv 0$), whereas at the ``free'' water surface the slip condition ($\nabla u \vec{e_z} | _{z=D}=\nabla v \vec{e_z} | _{z=D}\equiv 0$) and $w|_{z=D}\equiv 0$ were applied ($u$, $v$, and $w$ are the 
azimuthal, radial and vertical velocity components).

The numerical model is based on a finite-volume discretization of the governing equations on a regular cylindrical grid. 
The subgrid-scale turbulence is implicitly parameterized by the {Adaptive Local Deconvolution Method} (ALDM), see \cite{hickel_06}. Time integration is done using the explicit low-storage third-order Runge-Kutta method of 
\cite{williamson_1980}.

Three series of numerical simulations have been performed by cylFloit: the ``from scratch'' series (i), where the studied state at a target rotation rate $\Omega$ was reached after initializing the system from a non-rotating axially symmetric initial state; and the ``spin-up'' (ii) and ``spin-down'' (iii) series, where a rotation rate evolution $\Omega(t)$ similar to the aforementioned laboratory sequences was imitated. The numerical parameters of these simulations are listed in table \ref{cylFloit_grids}.

\paragraph{(i) The ``from scratch'' simulations:}  
In this initialization procedure, firstly an axially symmetric (thus, two dimensional; 2d) stationary solution was computed within a physical time of $t_{2\mathrm{d}}=10800\,\mathrm{s}\,(3\,\mathrm{hrs})$, with $\Omega=0$, but with the aforementioned boundary conditions. To obtain an axially symmetric solution, the number of azimuthal grid cells was set to $N_\theta=1$, thus reducing the problem to 2d. 
Then, starting from this state the full 3d simulation was initialized with a spin-up from zero 
angular velocity to its final value $\Omega_f$ as: 
	\begin{align}
	\label{eqn:validation_simulation_cylFloit_I}	
	&\Omega \left( t\right) = \begin{cases}
	0, & 0\le t \le t_{2\mathrm{d}}\\
	\frac{\Omega_{f}}{2}\{ 1 - \cos \left[ \frac{\pi}{\tau}
	\left(t - t_{2\mathrm{d}} \right)\right] \}  \\,
	t_{2\mathrm{D}} < t \le t_{2\mathrm{d}}+\tau \\
	\Omega_{f} , & t> t_{2\mathrm{d}}+\tau
	\end{cases}.
 	\end{align}
Here $\Omega_{f}$ is the final constant angular velocity used in the experiment and $\tau$ denotes the spin-up period 
of the rotating annulus 
ranging from $20$ s for $\Omega_{\rm min}$ to $910$ s for $\Omega_{\rm max}$ \citep{borchert}. 
To trigger the formation of baroclinic waves, low amplitude random perturbations were added to the temperature field, with a maximum amplitude of $\delta T_{\mathrm{pert}}=0.03 |T_b-T_a|$.  
This 3d simulation took further $10800$ s, so that the waves could fully develop.
A subsequent integration time of $7200$ s (2 hrs) at maximum was used to record the data analysed in the present work. 
For further information on this initialization method, we refer to \cite{borchert}.
 
\paragraph{(ii) Spin-up simulations:}
In these cases an initial angular velocity $\Omega_i$ and a final angular velocity $\Omega_f>\Omega_i$ were chosen. 
The time evolution of the angular velocity $\Omega(t)$ was then computed according to the formula: 
	\begin{align}
	\label{eqn:validation_simulation_cylFloit_I}	
	&\Omega \left( t\right) = \begin{cases}
	\Omega_i+\frac{\Omega_{f}-\Omega_i}{2}\left\{ 1 - \cos \left(\pi \frac{t}{\tau'}\right) \right\}  ,
	& t \leq \tau' \\
	\Omega_{f} , & t> \tau'
	\end{cases},
 	\end{align}
where $\tau'$ means the spin-up or spin-down period. 
The first simulation started with $\Omega_i = 0$ rpm and $\Omega_f = 2$ rpm, the second simulation used 
$\Omega_i = 2$ rpm and $\Omega_f = 3$ rpm, the third $\Omega_i = 3$ rpm and $\Omega_f = 4$ rpm and so forth 
up to the last spin-up simulation with $\Omega_i = 19$ rpm and $\Omega_f = 20$ rpm. 
The spin-up period was set to $\tau' = 1200$ s (20 min) in order to imitate the typical spin-up time scale of the laboratory runs. After the spin-up period the simulation took  
further $1800$s (30 min).  
Each simulation was initialized with fields from the previous simulation.

\paragraph{(iii) Spin-down simulations:}
The parameters and the procedures of the spin-down series were the same as for the spin-up runs, the only difference being that in this case $\Omega_i>\Omega_f$ holds.
The first spin-down simulation was initialized with the results from the last spin-up simulation. 
It therefore used $\Omega_i = 20$ rpm and $\Omega_f = 19$ rpm, the next $\Omega_i = 19$ rpm and $\Omega_f = 18$ rpm, 
and so forth down to $\Omega_i = 2$ rpm and $\Omega_f = 0$ rpm. 
After the spin-down period of  $\tau' = 1200$ s the simulations here took 
further $1200$ s only, which was long enough for the flow to equilibrate. 

\linespread{1.5}
\begin{table}
\begin{center}
 \begin{tabular}{lcc}\hline \hline
     & ``from scratch'' 	& ``spin-up/spin-down'' 	\\ \hline \hline
Points in $r$-$\theta$-$z$	& \multicolumn{2}{c}{$40 \times 60 \times 50$}	 \\ \hline
Points total		& \multicolumn{2}{c}{120000} \\ \hline
$\Delta r$	& \multicolumn{2}{c}{1.88 mm} \\ \hline
$(r \Delta \theta)_{\rm inner/middle/outer}$	& \multicolumn{2}{c}{4.71 / 8.64 / 12.57 mm} \\ \hline
$\Delta z$	& \multicolumn{2}{c}{2.7 mm} \\ \hline
Integration time step $\delta t$  & \multicolumn{2}{c}{dynamically adapted to satisfy} \\ [-0.5em]
 & \multicolumn{2}{c}{CFL-criterion$^1$ (mean value $\langle\delta t\rangle\approx 0.1$ s)} \\ \hline
Time between  & & \\[-0.5em]
two data outputs $\Delta t$  & 5 s &  3 s	\\ \hline 
 \end{tabular}
\linespread{1.0}
 \caption{Numerical parameters for the simulations with cylFloit. 
 ($^1$ Courant-Friedrichs-Lewy criterion.)}
 \label{cylFloit_grids}
\end{center}

\end{table}

\subsubsection{EULAG}
The EULAG framework is a multipurpose multi scale solver for all-scale geophysical flows, see \cite{Prusa2008} for a comprehensive review. The framework formulates the non-hydrostatic anelastic fluid equations of motion, e.g., \cite{Grabowski2002}, that can be solved either in Eulerian flux form or in semi-Lagrangian advective form, and it allows for a number of assumptions for particular flow characteristics, specifically the compressible/incompressible Boussinesq approximation, incompressible Euler/Navier-Stokes equations, and fully compressible Euler equations. The governing partial differential equations are evaluated with a semi-implicit non-oscillatory forward-in-time (NFT) algorithm and a finite volume discretization \citep{Smolarkiewicz1991,Smolarkiewicz1997,Smolarkiewicz1998}. EULAG has been successfully applied to a number of geophysical problems, documented by the large number of publications in the past years, see the list of publications with respect to applications on the EULAG model website at \url{http://www.mmm.ucar.edu/eulag/pub_appl.html}, ranging from cloud microscale to synoptic and global scale in atmospheric flows, as well as it was used for modeling oceanic flows. It is worth mentioning that also solar convection \citep{Elliott2002}, and urban flows \citep{Schroettle2013}, were studied, and beyond geoscience phenomena, EULAG has been also applied for simulating waves in the human brain \citep{Cotter2002}. Apart from the possibility to consider particular flow characteristics as mentioned above, EULAG also provides a framework for Direct Numerical Simulation (DNS), Large Eddy Simulation (LES), and implicit LES (ILES). We here use the DNS approach.

We adapted the general EULAG framework for our purposes. The sidewalls and the end walls of the annulus were modeled with the immersed boundary approach (cf.~\cite{Goldstein1993}), where fictitious body forces in the governing equation of motion are incorporated to represent no-slip boundaries which leads to a damping of the solution in an appropriate time interval. \cite{Smolarkiewicz2007} gives a detailed description of the implementation of the immersed boundary approach in the EULAG flow solver. In our study, the damping parameters were set so that the motion at the boundaries was damped to zero within a single time step. The time step increment was set to $\delta t=2.5\times 10^{-3}$ s. 

We used a Cartesian ($x,y,z$) domain with physical lengths $0.258\times0.258$ m in $x$- and $y$-direction and $0.135$ m in $z$-direction where $z$ is the height. The grid dimensions were $192\times 192\times 67$ cells in $x,y,z$. That results in a grid resolution of $\Delta x = \Delta y = 0.135$ m in $x,y$-direction and $\Delta z = 0.204\times 10^{-2}$ m in $z$-direction. The properties of the grid and the timestep are summarized in Table \ref{eulag_table}.

The temperature difference $\Delta T\equiv T_b - T_a$, with $T_b= T_{\rm ref}+\Delta T/2$ and $T_a= T_{\rm ref}-\Delta T/2$ is realized by setting $T=T_{b}\,(T_{a})$  where the radius is equal or greater (lower) than the outer (inner) radius. The reference temperature was set to $T_{\rm ref}=20^\circ$C and thus the temperature difference between the inner and outer layer was $\Delta T=8$ K, in agreement with the laboratory experiment. The governing equations (\ref{eq:momentum}) to (\ref{equ:temperature}) were solved in the Boussinesq approximation, and the centrifugal term $\Omega^2 r$ was also incorporated into (\ref{eq:momentum}), as in the case of cylFloit. For the $\rho(T)$ dependence a linear decrease of density with respect to temperature was assumed with volumetric thermal expansion coefficient $\alpha=2.07\times 10^{-4}$ K$^{-1}$, as given at $T_{\rm ref}$. The Prandtl number was set to $Pr=7$, corresponding to the properties of de-ionized water.

\linespread{1.5}
\begin{table}
\begin{center}
 \begin{tabular}{lc}\hline \hline
     & EULAG simulations 	\\ \hline \hline
Points in $x$-$y$-$z$	& {$192 \times 192 \times 67$}	 \\ \hline
Points total		& {$2\,469\,888$} \\ \hline
$\Delta x=\Delta y$	& $1.35$ mm \\ \hline
$\Delta z$	& $2.04$ mm \\ \hline
Integration time step $\delta t$  & $2.5\times 10^{-3}$ s \\ \hline
Time between  & \\[-0.5em]
two data outputs $\Delta t$  & 5 s	\\ \hline 
 \end{tabular}
\linespread{1.0}
 \caption{Numerical parameters for the simulations with EULAG.}
 \label{eulag_table}
\end{center}

\end{table}

\subsubsection{HiFlow$^3$}

HiFlow$^3$ is a multi-purpose C++ finite element software providing tools for efficient and accurate solution of a wide range of problems modeled by partial differential equations (PDEs), cf. \cite{2010Heuveline_HiFlow3,emcl-preprint-2012-05}.
It follows a modular and generic approach for building efficient parallel numerical solvers and introduces parallelity on two levels: coarse-grained parallelism by means of distributed grids and distributed data structures, and fine-grained parallelism by means of platform-optimized linear algebra back-ends (e.g. GPU, Multicore, Cell, etc.). 
Further information about this open source project can be found on the project's website \url{http://hiflow3.org/}.
For the baroclinic wave tank scenario the governing equations (\ref{eq:momentum}) to (\ref{equ:temperature}) were considered in cylindrical coordinates in a non-rotating frame, thus the Coriolis term of (\ref{eq:momentum}) was not present in this implementation.
The rotation of the system was hence taken into account by setting the proper boundary conditions at the lateral and bottom sidewalls for the azimuthal velocity component, corresponding
to the rigid body rotation of the cylinder:

\begin{equation}
\vec{u}|_{r=a}=\vec{u}|_{r=b}=\vec{u}|_{z=0} = r\Omega\vec{e}_{\theta},\label{equ:bc_vel}
\end{equation}
where $\vec{e}_{\theta}$ denotes the unit vector in the azimuthal
direction. The rest of the boundary conditions were of identical types to those described in the case of the cylFloit model, but the actual values of the sidewall temperatures were set differently, as: $T_a=20^\circ$C and $T_b=28^\circ$C.
Material parameters $\nu$ and $\kappa$ were set constant (with their standard values for de-ionized water at reference temperature $T_{\rm ref}=20^\circ$C), and for the thermal expansion the linear form of 
\begin{equation}
\frac{\delta\rho}{\rho_0}=-\alpha(T-T_{\rm ref}),
\end{equation}
was used with the standard value of $\alpha=2.07\times 10^{-4}$ K$^{-1}$. 

For the calculation of the initial temperature and velocity fields
the stationary version of the governing equations (\ref{eq:momentum})-(\ref{equ:temperature}) were considered,
i.e:
\begin{eqnarray}
(\vec{u}\cdot\nabla)\vec{u} & = & -\frac{1}{\rho_{0}}\nabla p-\alpha(T-T_0)g\vec{e}_{z}+\nu_{i}\nabla^2\vec{u},\label{eq:momentum-ini}\\
\nabla\cdot\vec{u} & = & 0,\label{equ:continuity-ini}\\
(\vec{u}\cdot\nabla)T & = & \kappa_{i}\nabla^2 T,\label{equ:temperature-ini}
\end{eqnarray}
with the aforementioned boundary conditions. For the initialization phase increased values of thermal diffusivity and kinematic viscosity were used, in the forms of $\nu_{i} = 100\cdot\nu$ and
$\kappa_{i} = 100\cdot\kappa$ in order to determine stationary solutions.
Since the rotation was already implemented as a
boundary condition of the stationary problem -- that served as initial condition for the time-dependent simulation to follow -- no spin-up or spin-down was applied. 

In some of the runs temperature perturbations
were also added to the initial stationary temperature fields to reveal whether these perturbations
have an influence on the developing baroclinic wave patterns. This temperature perturbation
is defined in terms of the maximum perturbation $M$ [K] and azimuthal wave number $k$:
\[
\delta T(r,\theta,z)=M\sin\left(\frac{r-a}{b-a}\pi\right)\cos(k\theta)\sin\left(\frac{z}{D}\pi\right),
\]
for $r\in[a,b]$, $\theta\in[0,2\pi]$, and $z\in[0,D]$.
In the perturbed numerical simulations presented in this paper, $M=0.25\,\text{K}$ and $k=1$ was chosen.

The resulting stationary velocity field and the corresponding (occasionally
perturbed) temperature field were used as the initial conditions $\vec{u}_{0}$
and $T_{0}$ of the time-dependent problem and a simulation time of between about 
$1,000\,\text{s}$ up to $2,500\,\text{s}$ have been considered. The governing equations 
are solved on a cylindrical mesh with $65,436$ points based on a finite element 
method. Cellwise tri-quadratic velocity and temperature functions 
and piecewise tri-linear pressure functions were used. This type of so-called Taylor-Hood
elements are known to be stable in the sense that they fulfil the inf-sup condition \citep{1974Brezzi}. In Table 
\ref{tab:grid_hiflow3}, an overview of the grid and the points of the degrees of 
freedom (DOF) of the applied finite element method is given. On this grid, the 
state of the discrete solution (velocity, pressure, and temperature) is described 
by $N=2,084,604$ DOF at each point in time.

\begin{table*}
\begin{centering}
\begin{tabular}{ccc}
\hline 
 & \textbf{Grid points / DOF points for pressure } & \textbf{DOF points for velocity and temperature}\tabularnewline
\hline \hline 
Points in $r-\theta-z$ & $21\times76\times41$ & $41\times152\times81$\tabularnewline
\hline 
Points total & $65,436$ & $504,792$\tabularnewline
\hline 
$\Delta r_{\text{min}/\text{max}}$ & $2.785/5.250\,\text{mm}$ & $1.438/2.625\,\text{mm}$\tabularnewline
\hline 
$r\Delta\theta_{\text{min}/\text{middle}/\text{max}}$ & $3.720/6.821/9.921\,\text{mm}$ & $1.860/3.410/4.961\,\text{mm}$\tabularnewline
\hline 
$\Delta z_{\text{min}/\text{max}}$ & $1.700/5.625\,\text{mm}$ & $0.850/2.813\,\text{mm}$\tabularnewline
\hline 
Integration time step $\delta t$ & $0.25$ s & $0.25$ s \tabularnewline
\hline 
Time between two data outputs $\Delta t$ & $0.25$ s & $0.25$ s \tabularnewline
\hline 
\end{tabular}\caption{Parameter overview of the grid and the Lagrange
points (DOF) of the applied finite element method by HiFlow$^3$.}
\label{tab:grid_hiflow3}
\par\end{centering}
\end{table*}

%
In time, the Crank-Nicholson scheme was applied to (\ref{eq:momentum})-(\ref{equ:temperature})
resulting in a fully coupled nonlinear equation system with all $N$ unknowns
for each time step. The time step (as well as the data output interval) was 
set to $\Delta t =\delta t=0.25\,\text{s}$.

For the solution of the nonlinear problem in each time step,
Newton's method was applied. In a typical time step, 2 or 3 steps of the Newton
iteration were sufficient to solve the problem adequately. The linear
equation system within each Newton step is assembled and solved on
a High-Performance Computer system.
A GMRES solver has been applied with block-wise incomplete LU preconditioner 
(ILU++ \cite{2007ILUPP}), which required ca. 200 iterations in a typical calculation.

\subsubsection{INCA}
INCA is a multi-purpose engineering flow solver for both compressible and incompressible problems using Cartesian adaptive grids and an immersed boundary method to represent solid walls that are not aligned with grid lines. INCA has successfully been applied to a wide range of different flow problems, ranging from incompressible boundary layer flows \citep{hickel_08} to supersonic flows \citep{grilli}.

In the current context the incompressible module of INCA was used with an extension to fluids with small density perturbations governed by the Boussinesq equations (\ref{eq:momentum}) to (\ref{equ:temperature}) in a co-rotating reference frame. The governing equations are discretized by a finite-volume fractional-step method \citep{chorin} on staggered Cartesian mesh blocks. For the spatial discretization of the advective terms the Adaptive Local Deconvolution Method (ALDM) with implicit turbulence parameterization was used \citep{hickel_06}. For the diffusive terms and the pressure Poisson solver a non-dissipative central scheme with 2nd order accuracy was chosen. For time advancement the explicit third-order Runge-Kutta scheme of \cite{shu} was used. The time step is dynamically adapted to satisfy a Courant-Friedrichs-Lewy condition with $CF L \leq 1.0$. The Poisson equation for the pressure is solved at every Runge-Kutta sub-step, using a Krylov subspace solver with algebraic-multigrid preconditioning.
The general applicability of INCA in the Boussinesq approximation with ALDM as an implicit turbulence SGS model to stably stratified turbulent flows has been demonstrated in \cite{rh_12} and \cite{rh_13}. 

To represent the annulus geometry within Cartesian grid blocks in INCA, two cylindrical immersed boundaries were used representing the lateral sidewalls of the flow cavity. The Conservative Immersed Interface Method of \cite{meyer} was employed to impose the boundary condition, that were of the same types as described in the subsection for cylFloit. The sidewall temperatures were $T_a=16^\circ$ C and $T_b=24^\circ$ C (yielding $\Delta T = 8$ K), and the density changes with temperature were parametrized in a linear approximation, with the same value of $\alpha$ as for EULAG and HiFlow$^3$.

For the simulations presented here a grid with $160\times 160 \times 90$ cells was used ($x$, $y$ and $z$ direction, $z$ being the vertical), i. e. 2.3 million cells. The grid was equidistant in the horizontal directions and refined towards the bottom wall in the vertical direction. The domain was split into 32 grid blocks for parallel computing. 

The simulations were initialized with a stable temperature stratification. At $t=0$ the wall temperature and the rotation were switched on instantaneously (no spin-up or spin-down was applied).
As mentioned above, during the initial phase, the integration time step was adjusted dynamically and fluctuated around $\delta t \approx 0.05$ s. In the period of constant step size, its value was $\delta t = 0.0375$ s. 
The total physical duration of each run ranged from $750$ s to $1500$ s, and the output time step was set to $\Delta t = 5.625$ s. The numerical parameters of the INCA simulations are summarized in table \ref{inca_table}.

\linespread{1.5}
\begin{table}
\begin{center}
 \begin{tabular}{lc}\hline \hline
     & INCA simulations 	\\ \hline \hline
Points in $x$-$y$-$z$	& {$160 \times 160 \times 90$}	 \\ \hline
Points total		& {$2\,304\,000$} \\ \hline
$\Delta x=\Delta y$	& $1.55$ mm \\ \hline
$\Delta z_{\rm min/max}$	& $0.4\,/\,1.8$ mm \\ \hline
Integration time step $\delta t$ & \\ 
in the initial phase  & $\approx 0.05$ s \\ \hline
Integration time step $\delta t$ & \\ 
in the constant step phase  & $0.0375$ s \\ \hline
Time between  & \\[-0.5em]
two data outputs $\Delta t$  & $5.625$ s	\\ \hline 
 \end{tabular}
\linespread{1.0}
 \caption{Numerical parameters for the simulations with INCA.}
 \label{inca_table}
\end{center}

\end{table}

\subsubsection{LESOCC2}

The multi-purpose solver LESOCC2 \citep{froehlich2006,Hinterberger2007} was used to solve the governing equations (\ref{eq:momentum}) to (\ref{equ:temperature}) in cartesian coordinates from a co-rotating reference frame.
The discretization method applied is a finite volume method with a collocated variable arrangement on curvilinear coordinates. For time integration a fractional step method was employed, consisting of a Runge-Kutta scheme as predictor and a pressure-correction equation as corrector \citep{Zhu1992b}. The momentum interpolation of \cite{Rhie1983} was incorporated in the discretization for pressure-velocity coupling. Parallelization was realized by domain decomposition on the basis of block-structured grids and was implemented with MPI.

Similarly to cylFloit (and to the actual experiment) ``spin-up'' and ``spin-down'' sequences were conducted.
The first simulation of the ``spin-up'' sequence was initiated from a stably stratified axially symmetric, non-rotating state. Then the rotation was switched on immediately. The next simulation at a higher rotation rate $\Omega$ was initiated analogously, but this time the final velocity and temperature fields of the preceding simulation were used as initial conditions. This procedure was repeated until $\Omega_{\rm max}=20$ rpm was reached (in 8 subsequent simulations), and then the backward (``spin-down'') series started, in which the runs were initiated from the final state obtained at a higher $\Omega$, in the same manner.

The boundary conditions had the same type as for cylFloit or INCA, the only difference being the temperatures prescribed at the lateral sidewalls, which in the case of LESOCC2 were $T_a=23.5^\circ$C and $T_b=31.5^\circ$C. The reference temperature and the thermal expansion coefficient $\alpha$ were set as discribed for the HiFlow$^3$ simulations.
For discretization two different non-equidistant grid meshes were used, whose properties are listed in Table \ref{dd_grids}.
The grids employed were curvilinear, body-fitted and block structured.
The time steps were adapted automatically due to a combined convection-diffusion criterion, and varied in the regime: $\delta t \in (0.0177 \,{\rm s}; 0.0377\, {\rm s})$.
The data output time step was set to $\Delta t=1$ s.

\linespread{1.5}
\begin{table*}
\begin{center}
\begin{tabular}{lll} \hline \hline
    & \textbf{Grid 1}   & \textbf{Grid 2}       \\ \hline \hline
Points in $r$-$\theta$-$z$      & $76 \times 213 \times 137$    & $86 \times 241 \times 153$ \\ \hline
Points total            & 2.218 M       & 3.171 M \\ \hline
Blocks in $r$-$\theta$-$z$ (CPUs)       & $ 2 \times 4 \times 4$ (32)   & $2 \times 4 \times 8$ (64) \\ \hline
$\Delta r_{\rm min/max}$        & 0.6 / 1.4 mm  & 0.4 / 1.6 mm \\ \hline
$(r \Delta \theta)_{\rm inner/middle/outer}$    & 1.3 / 2.45 / 3.5 mm \phantom{mm}      & 1.2 / 2.2 / 3.1 mm \\ \hline
$\Delta z_{\rm min/max}$        & 0.6 / 1.1 mm  & 0.3 / 1.0 mm \\ \hline
average $\delta t$        & $\approx 0.033$ s  & $\approx 0.018$ s \\ \hline
Used for Simulation with        & $\Omega < 17$ rpm     & $\Omega\geq 17$ rpm   \\ \hline
 \end{tabular}
 \caption{Overview of grids used for baroclinic wave simulations by LESOCC2.}
 \label{dd_grids}
\end{center}

\end{table*}
\linespread{1.0}
\subsection{Data processing}
\begin{figure}
\noindent\includegraphics[width=8cm]{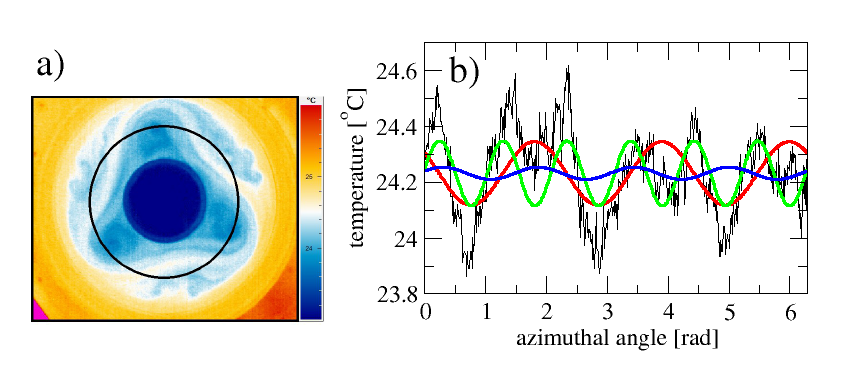}
 \centering
\caption{Three steps of data processing, demonstrated on a single thermographic snapshot of the laboratory experiment.
The temperature values of the raw image are (a) extracted along a circular contour at mid-radius $r_{\rm mid}$, thus the azimuthal temperature profile (b) is obtained. The Fourier components of integer wave numbers are then determined for each time step. In this exemplary case modes $m=3,4$ and 6 are shown by red, blue and green curves, respectively.}
\label{dataproc}
\end{figure}

To reduce the parameter space to investigate, from the (either experimentally or numerically) obtained temperature fields close to the free water surface a path-wise temperature profile $T(\theta)$ was extracted along a circular contour at mid-radius $r_{\rm mid}=(a+b)/2=8.25$ cm for each available time instant (black circle in the exemplary experimental thermographic image in Fig. \ref{dataproc}a). In the cases where the temperature data were stored in Cartesian grids (i.e. for INCA and for the laboratory experiment itself), linear interpolation was applied to gain equally spaced azimuthal temperature profiles (e.g. the black curve of Fig. \ref{dataproc}b). During post-processing the data were transformed so that the azimuthal angle $\theta$ was measured clockwise from a given co-rotating point. For the experimental and HiFlow$^3$ data -- which were given in the reference frame of the laboratory -- the rotation of the tank also had to be compensated to yield the appropriate co-rotating measure of $\theta$.       

As mentioned before, the experimentally observed thermal structures were considered surface ($z=D=13.5$ cm) temperature patterns. Also in the cases of EULAG, HiFlow$^3$ and LESOCC2 the temperature fields of the uppermost grid level were considered.
For cylFloit and INCA, however, the temperature profiles were extracted from the somewhat lower level of $z=10$ cm.

In order to determine the dominant azimuthal wave modes and their corresponding amplitudes and drift rates (to be discussed in the next section), the temperature profiles $T(t,\theta)$ were analyzed using discrete spatial Fourier decomposition. After subtracting the mean temperature $\langle T(\theta ; t)\rangle$ (averaged over the whole azimuthal domain of the contour at each time instant $t$), the remaining fluctuations could be expressed as amplitudes $A_m(t)$ and phases $\phi_m(t)$ of trigonometric functions with integer wave numbers $m=1,2,\cdots$, as:
\begin{equation}
T(\theta ; t)-\langle T(\theta ; t)\rangle \approx \sum_{m}A_m(t)\, \sin(m \theta + \phi_m(t)).
\label{decomp}
\end{equation} 
Fig. \ref{dataproc}b demonstrates this step, showing three (exemplarily selected) components: $m=3$ (red), $m=4$ (blue) and $m=6$ (green) at a given time instant. The time series of $A_m(t)$ and $\phi_m(t)$ of the different numerical models and the experiments could then be easily compared using various standard methods of signal processing, to be discussed in the following section.

\section{Results}
\subsection{Wave numbers}

Firstly, the time averaged amplitudes $\langle A_m \rangle$ of the spatial Fourier components were determined in each (either experimental or numerical) case using the above described methodology. For this averaging the transient part of the wave evolution was 
omitted, only the quasi-stationary part of each time series was retained. 

These time averaged spatial Fourier spectra showed, that besides the wave number corresponding to the main azimuthal symmetry properties of a given baroclinic wave, the smaller-scale structures of the surface temperature field also leave a pronounced spectral ``fingerprint''. In the Fourier space, these patterns are represented as harmonics of the basic wave number. It is to be emphasized, that the term `harmonic' here is meant strictly in the sense of integer multiples of the wave number, without any further implications on the dynamics. 

\subsubsection{A conceptual demonstration}

\begin{figure}
\noindent\includegraphics[width=8cm]{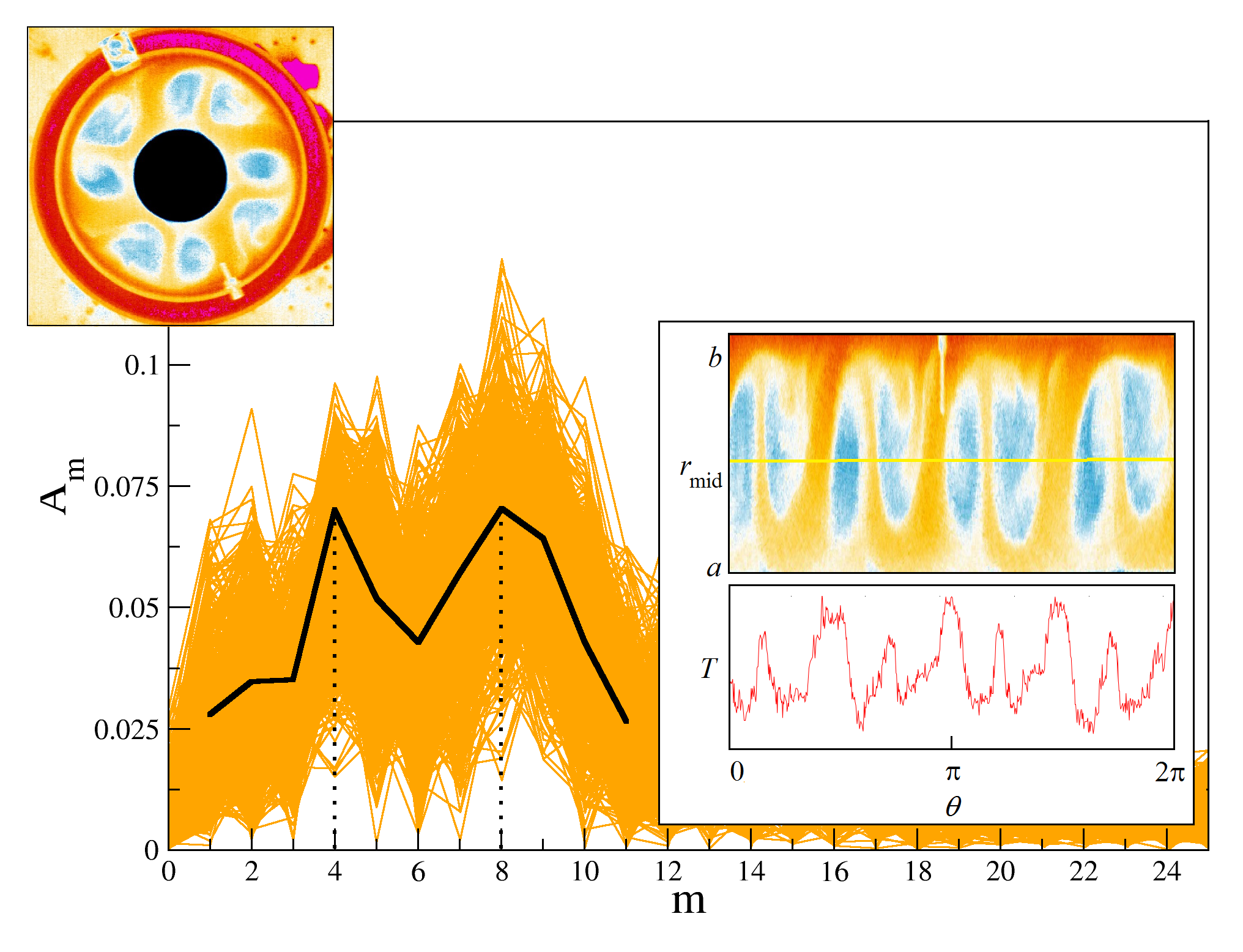}
 \centering
\caption{Spatial Fourier spectra (orange), extracted from the quasi-stationary part of a laboratory experiment ($\Omega=17.1$ rpm, spin-down series), and their temporal average in the lower $m$-domain (black). In the inset, a typical thermographic snapshot is shown in polar coordinates, and the corresponding one dimensional temperature profile at $r_{\rm mid}$ (red curve).}
\label{spectra}
\end{figure}

As a demonstration of the physical origin of such spectral peaks, an exemplary case is shown in Fig. \ref{spectra}. The top left inset shows one of the original images of a given laboratory experiment, where the four-fold symmetric shape of the temperature field is apparent. The bottom right inset depicts the same image as transformed to polar coordinates: the yellow line marks mid-radius $r_{\rm mid}$, and the corresponding pathwise temperature profile is also given underneath. The spatial Fourier spectra of such profiles, taken at different time instants during the same experimental run are plotted as orange curves in the main panel. Their average is also indicated (thick black curve). Manifestly, alongside the peak of $m=4$, another significant spectral peak appears at $m=8$, caused by the warm jet that is meandering between cold eddies (cf. insets). 

In several cases among the laboratory experiments, such geometric ``harmonics'' even surpassed the ``basic mode'' in amplitude. Therefore, in order to be consistent with the traditional visual classification of wave numbers, not necessarily the largest peak was labeled the so-called \emph{dominant wave number}. Instead, the following algorithm was applied: (i) all the significant peaks of the time-averaged spectra were determined. (ii) If two or more peaks appeared at wave numbers that are integer multiples of the first one, then the wave number $m$ of the first peak was considered to be the dominant wave number. Even if its average amplitude $\langle A_m \rangle$ is not the largest of all, this definition still implies that the patterns bear an overall symmetry to azimuthal rotation by $2\pi/m$ (i.e. the autocorrelation of the temperature profile exhibits its largest positive peak at $2\pi/m$).

\subsubsection{The dominant wave numbers}
   
\begin{figure}
\noindent\includegraphics[width=8cm]{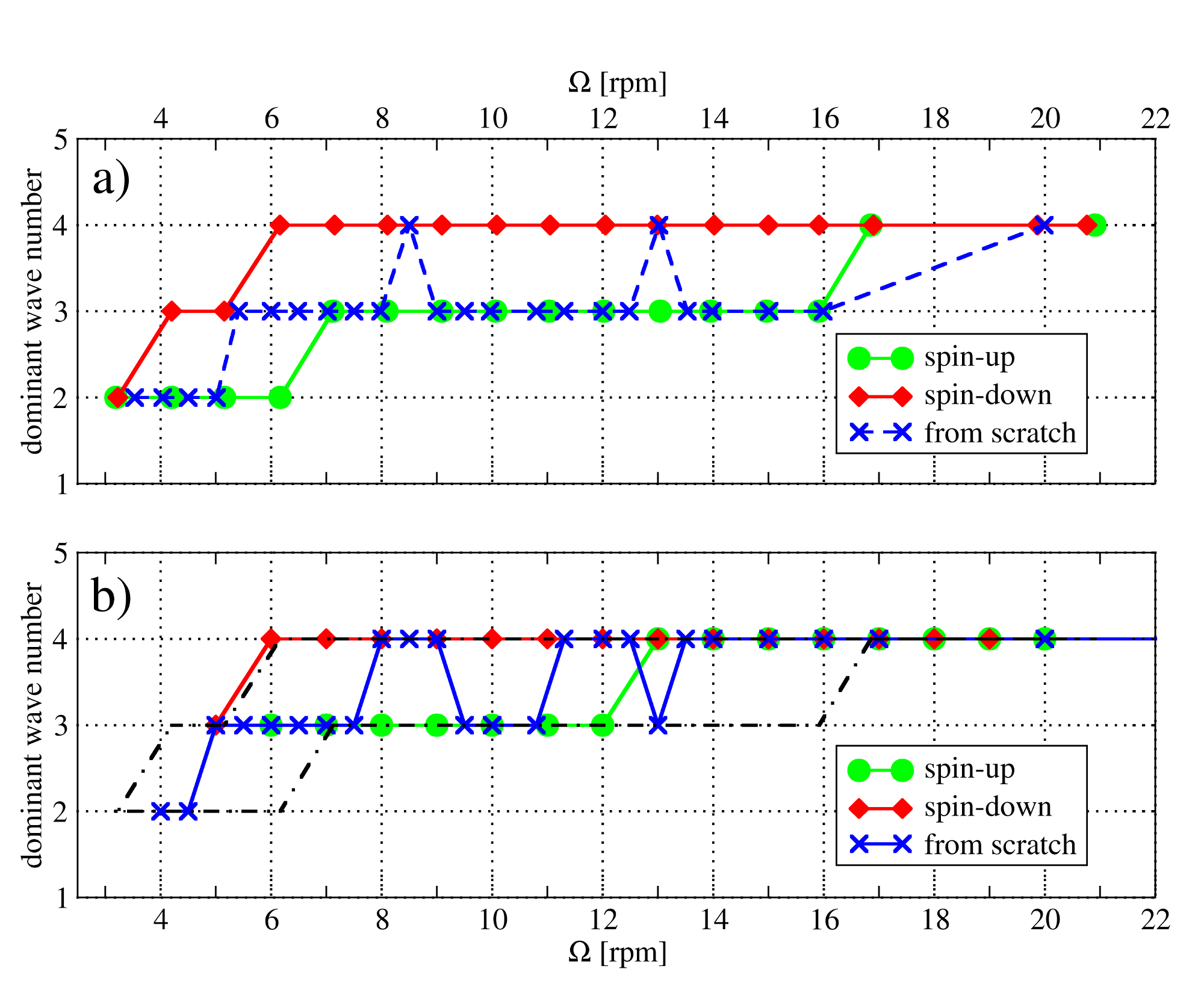}
 \centering
\caption{``Subway map'' of the baroclinic annulus: the dominant wave numbers as a function of rotation rate $\Omega$ as found in the experiments (a) and in the cylFloit simulations (b).}
\label{subway}
\end{figure}

The above defined dominant wave numbers are presented in Fig. \ref{subway}a as a function of rotation rate $\Omega$, as found in the laboratory experiments.
Apparently, large hysteresis can be observed (in qualitative agreement with the findings of several previous studies \citep{miller_butler,christoph_book,thomas_npg}), implying multiple equilibria.
A broad rotation rate regime (ranging from 3.9 rpm $< \Omega <$ 17.1 rpm) exhibited different wave numbers in the ``spin-up'' and ``spin-down'' series, with $m=3$ and $m=4$ being the dominant modes, respectively (see green and red curves in Fig. \ref{subway}). It is important to note that even in the hysteretic regime the wave 
patterns appeared to be stable against surface perturbations: during the experimentation process, after recording 
a particular pattern, irregular manual stirring was applied in the uppermost fluid layer (with penetration depth 
of roughly $1$ cm), and afterwards, in all observed cases, the same wave pattern recovered within ca. 10 minutes of time. Despite the hysteresis, it is to be remarked, that the critical rotation rate $\Omega_{\rm crit}\approx 3$ rpm of the onset of baroclinic 
instability and the first -- critical -- wave number ($m_{\rm crit}=2$) appeared to be unaffected by the initial conditions.
 
These marked phenomena motivated the numerical approach applied in the cylFloit and LESOCC2 runs, which imitated 
the experimental process via initiating the simulation of a given parameter point from the final flow state 
of the preceding simulation. By sequentially increasing (decreasing) the rotation rate in this manner, ``spin-up'' (``spin-down'') series were generated, as discussed in the previous section. 
Besides, the stability of the obtained states to perturbed initial states (the analogue of manual surface stirring in 
the laboratory) was analyzed in the HiFlow$^3$ simulations. 

The dominant wave numbers of the cylFloit runs are shown in Fig. \ref{subway}b. The green and red curves 
represent the spin-up and spin-down series, respectively. 
Compared to the experimental data of panel a), the cylFloit spin-up series exhibited switches from $m=2$ 
to $m=3$ and from $m=3$ to $m=4$ at lower rotation rates. Nevertheless, it can be stated, that throughout the 
whole series, the simulations always converged to one of the experimentally observed equilibria, i.e. the cylFloit 
spin-up curve is enveloped by the experimental hysteresis regime (repeated in Fig. \ref{subway}b with dash-dotted lines). The spin-down series, on the other hand, precisely reproduced the laboratory results, including 
the appropriate estimation of the critical rotation rate $\Omega_{\rm crit}$ and critical unstable mode $m_{\rm crit}$.

The dominant wave numbers obtained in an earlier experimental series (that was conducted in 2011 and had also been used for the validation of the cylFloit and INCA models, see the paper of \cite{borchert} in the present issue) are also shown in Fig. \ref{subway}a in the form of a blue curve. Each of these laboratory runs had been initiated with \emph{zero angular velocity} until the axially symmetric basic state of ``sideways convection'' developed. Afterwards, the rotation of the tank was accelerated so that the final rotation rate was reached within a spin-up period of ca. 20 s. In these experiments the wave patterns were observed -- and remained stable -- for extremely long times ranging from 6 to 12 hours after the onset of rotation.       
This laboratory procedure was also simulated with cylFloit (using the spin-up strategy described in the ``Numerical methods'' section), and the resulting data points are shown as the blue curve of Fig. \ref{subway}b. It can be stated that both in the experiments and in the simulations, even though the system was initiated ``from scratch'' before each run, the flow occasionally converged to the states of the upper (spin-down) branch. This observation underlines the conclusion that the hysteretic regime indeed involves two distinct equilibrium states (bifurcation) and does not arise merely due to some slow transient phenomenon. 

The experimental and numerical results for the four benchmark parameter points (for which the flow states were computed by all the numerical models) are summarized in Table \ref{mode_table}. These points were selected to represent the three dynamical regimes observed in the laboratory: the transition zone from axisymmetric ($m=0$) to wave flow state (\#1), the hysteretic regime (\#2 and \#3), and the regime of higher rotation rates, where -- at least in terms of the dominant wave numbers -- the two bifurcated branches have recombined (\#4). The arrows ($\uparrow$ and $\downarrow$) mark the spin-up and spin-down series, if applicable. In the case of the LESOCC2 runs, the flow states were also computed at intermediate data points (at rotation rates $\Omega=5.5;\,8.0;\,10.5;\,20.0$ rpm), to enable the same sequential simulation process as described for cylFloit. The data from these points, however, were not evaluated in the present study. 

\begin{table*}
    \begin{tabular}{ | l | l | l | l | l | l | l | l | p{5cm} |}
    \hline
    notation & $\Omega$ [rpm] & experiment & cylFloit & EULAG & HiFlow$^3$ & INCA & LESOCC2 \\ \hline
\#1 & $3\pm0.2$ & $0-2 (\uparrow \downarrow)$ & $0 (\uparrow \downarrow)$ &	$2-3I$ & $0(b,p)$ & $2$ & $0 (\uparrow \downarrow)$\\ \hline
\#2 & $7\pm0.1$ & $3(\uparrow);\,4(\downarrow)$ &	$3(\uparrow);\,4(\downarrow)$  & $3$ & $3(b);\,2(p)$ & $4$ & $2(\uparrow);\,3(\downarrow)$\\ \hline
\#3 & $9\pm0.1$ & $3(\uparrow);\,4(\downarrow)$  &	$3(\uparrow);\,4(\downarrow)$ 	& $4$ & $2(b);\,3(p)$ & $4$ & $3(\uparrow);\,4(\downarrow)$\\ \hline
\#4 & $17\pm0.1$ & $4 (\uparrow \downarrow)$  &	$4 (\uparrow \downarrow)$ &	$4$ & $4 (b,p)$ & $4$ & $3(\uparrow);\,4(\downarrow)$\\ \hline    
    \end{tabular}
    \caption{Dominant wave numbers of the ``benchmark'' data points, as obtained in the experiment and by the numerical models. Arrows $\uparrow$ and $\downarrow$ mark spin-up and spin-down initial conditions, if applicable. $b$ marks the basic and $p$ denotes the perturbed initiation states in the HiFlow$^3$ simulations. Note, that $\Delta T=8$ K was set constant for all the measurements, therefore the rotation rate $\Omega$ was the only variable ``environmental'' parameter.}
      \label{mode_table}
\end{table*}

In the case of the HiFlow$^3$ simulations, letters ``$b$'' and ``$p$'' denote the basic and perturbed states obtained for the given rotation rate, respectively.
In the ``$p$''-runs additional azimuthal random perturbation was added to the initial condition (described in the previous section). In the cases of \#2 and \#3, the perturbed initial state led to a solution different from the basic state, but no such behavior was found for \#1 and \#4. This is in qualitative agreement with the laboratory results, since all of these metastable states were found within the hysteretic regime. It is to be noted, that the LESOCC2 and HiFlow$^3$ models exhibited $m=2$ at \#2 in their spin-up and perturbed series, respectively, besides the (experimentally verified) $m=3$ mode.

EULAG and INCA always converged to one of the experimentally detected states within the regime of baroclinic instability (\#2 to \#4). For the data point \#1 close to the critical transition point, INCA found $m=2$, and EULAG showed an irregular pattern with fluctuating amplitudes at $m=2$ and $m=3$ (denoted with $2-3I$ in Table \ref{mode_table}), see also \cite{thomas_mz} in the present issue. These findings are seemingly in contradiction to the axially symmetric solutions of the rest of the models.
It is important to remark, however, that the exact experimental value of $\Omega_{\rm crit}$ is hard to determine. At $\Omega = 2.26$ rpm the flow in the laboratory tank was clearly axially symmetric, and at the next measured data point ($\Omega = 3.19$ rpm) the first baroclinic wave pattern with $m_{\rm crit}=2$ has already emerged. Moreover, in the aforementioned 2011 experimental series, axially symmetric ($m=0$) state was reported at $\Omega = 2.99$ rpm. Therefore the transition from $m=0$ to $m_{\rm crit}=2$ appears to take place at 3 rpm $< \Omega_{\rm crit} < $ 3.19 rpm, a rather narrow range. 

\subsubsection{Spatial harmonics and small-scale structure}
\begin{figure}
\noindent\includegraphics[width=8cm]{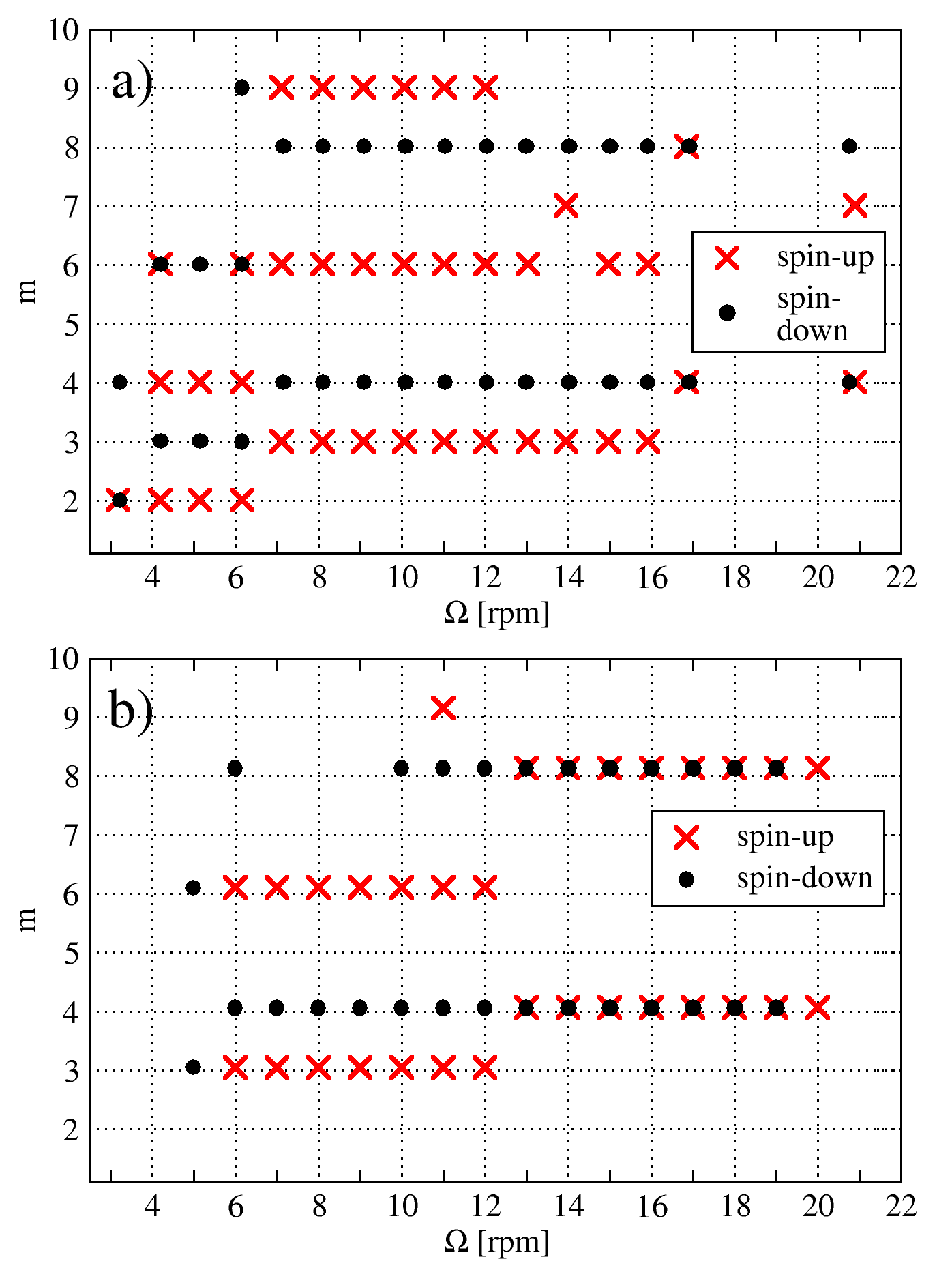}
 \centering
\caption{The distribution of significant harmonic modes in the wave number space, as a function of rotation rate $\Omega$, as found in the laboratory experiments (a) and in the cylFloit simulations (b).}
\label{harmonics}
\end{figure}

Besides the dominant wave numbers, the aforementioned ``harmonics'' are also of relevance, as they provide a certain spectral fingerprint of the studied patterns. 
The wave numbers corresponding to all significant peaks of the time-averaged spatial spectra are shown in Figs. \ref{harmonics}a and b, for the laboratory experiments and for the cylFloit runs, respectively. In both panels, red crosses mark the spin-up and black circles mark the spin-down series. 
In each case, a peak was considered significant if its time-averaged spatial Fourier amplitude $\langle A_m \rangle$ was larger than
$\bar{A}+3\sigma_A$, where $\bar{A}$ and $\sigma_A$ are the mean amplitude and standard deviation of the whole time-averaged spectrum, respectively.

Comparing the two panels of Fig. \ref{harmonics}, it is visible that in the laboratory experiments the presence of the harmonics was more pronounced than in the simulations. For example, the dominant wave mode $m=4$ was always accompanied by a significant $m=8$ in the laboratory (cf. Fig. \ref{spectra}), whereas it exhibited insignificant amplitudes in some of the cylFloit runs. Also, the harmonic $m=9$ regularly appeared alongside mode $m=3$ in the experimental data, whereas in the cylFloit results it showed up in one single case only.
This mismatch might indicate that the formation of some of the eddies in the annulus (that yield the presence of these harmonics) can be caused by surface effects (e.g. wind stress, nonzero heat flux, etc.) that are not included in the numerical models.  

\subsection{Average temperature variance}
  
\begin{figure*}
\noindent\includegraphics[width=14cm]{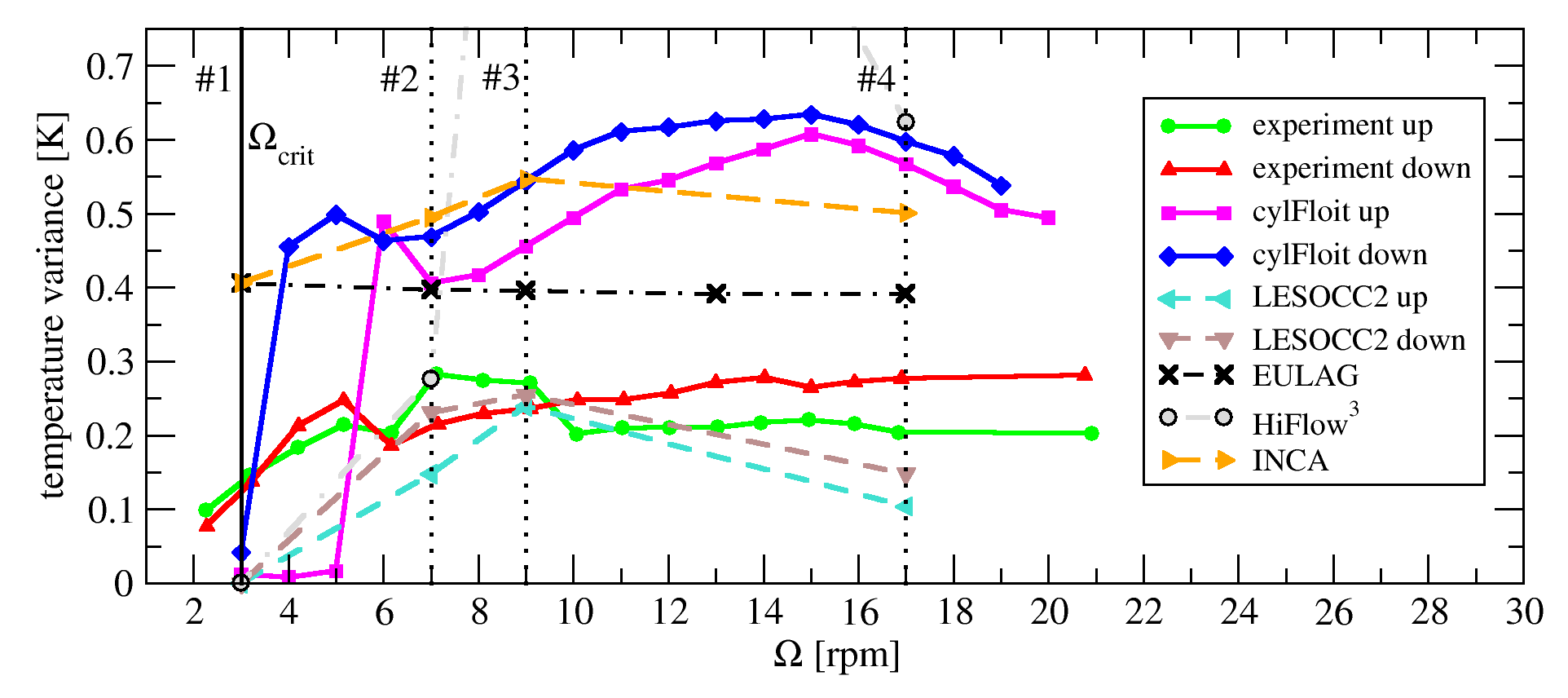}
 \centering
\caption{The average thermal variability $\bar{\sigma}$ as the function of rotation rate $\Omega$.}
\label{stdevs}
\end{figure*}

As a measure of the overall spatial thermal variability in the azimuthal direction, the (spatial) standard deviation of the mid-radius temperature profile was determined at each time instant. Next, the (temporal) average of these values -- denoted by $\bar{\sigma}$ -- was calculated for the whole quasi-stationary part of the given (either experimental or numerical) run.   
The obtained values are shown in Fig. \ref{stdevs} as a function of the rotation rate $\Omega$.
In the graphs corresponding to those numerical simulations, where the onset of baroclinic instability was captured, this ``phase transition'' manifests itself in the form of a marked jump in $\bar{\sigma}$. Note, that EULAG and INCA found dominant modes of non-zero $m$ already at the benchmark point \#1, therefore in their graphs no such jump is present. Since the basic state is axially symmetric and the analyzed data were extracted from a circular contour of a constant radius $r_{\rm mid}$, it is trivial that the numerical models give practically zero variance in this regime.
However, due to random temperature fluctuations, the laboratory experiments (green and red curves for the spin-up and spin-down series, respectively) showed considerably larger (yet, minimum) values of $\bar{\sigma}$ in this regime.

The qualitative behavior of the spin-down experimental series in terms of $\bar{\sigma}$ is well captured by the corresponding cylFloit runs (blue curve). In both curves pronounced local maxima can be observed at $\Omega = 5$ rpm, followed by local minima at $\Omega = 6$ rpm. Both in the experiments and the cylFloit runs, this parameter point coincides with the transition from dominant wave number $m = 4$ to $m=3$ (as we now discuss the spin-down sequence). This may imply that the $m=3$ patterns generally have larger amplitudes in the mid-radius section than their $m=4$ counterparts. Thus, the reorganization of the surface pattern overrides the general decreasing trend of $\bar{\sigma}$ towards smaller values of $\Omega$. A similar jump-wise increment is present in the experimental spin-up series as well (green curve). In this case, the transition happened at $\Omega=7$ rpm, which, again, coincides with the transition to $m=3$, this time from the preceding $m=2$ state (cf. Fig. \ref{subway}a). In this sequence also a similarly sharp drop of $\bar{\sigma}$ can be observed at $\Omega=10$ rpm, which is \emph{not} accompanied with the change of the dominant wave number $m=3$. However, as it will be demonstrated in the next subsection, this decrease coincides with a similarly sharp change in the drift rates of the baroclinic waves, thus implying a certain state transition, even though not in terms of $m$.

Despite the qualitative similarity, the cylFloit and INCA runs (blue, magenta and orange curves) systematically overestimate $\bar{\sigma}$ by around a factor of $2$.
This, however, may well be the consequence of the fact that the temperature fields of these models are extracted from the height level of $z=10$ cm (whereas $D=13.5$ cm).
The plotted data from LESOCC2 and EULAG (brown and black data points) on the other hand were extracted from the uppermost (surface) grid level. In terms of $\bar{\sigma}$, the former is in fairly good agreement with the experimental data, whereas the latter stays practically constant (exhibiting a minor decreasing trend only), and significantly overestimates the variance.

\subsection{Drift rates}

\begin{figure}
\noindent\includegraphics[width=8cm]{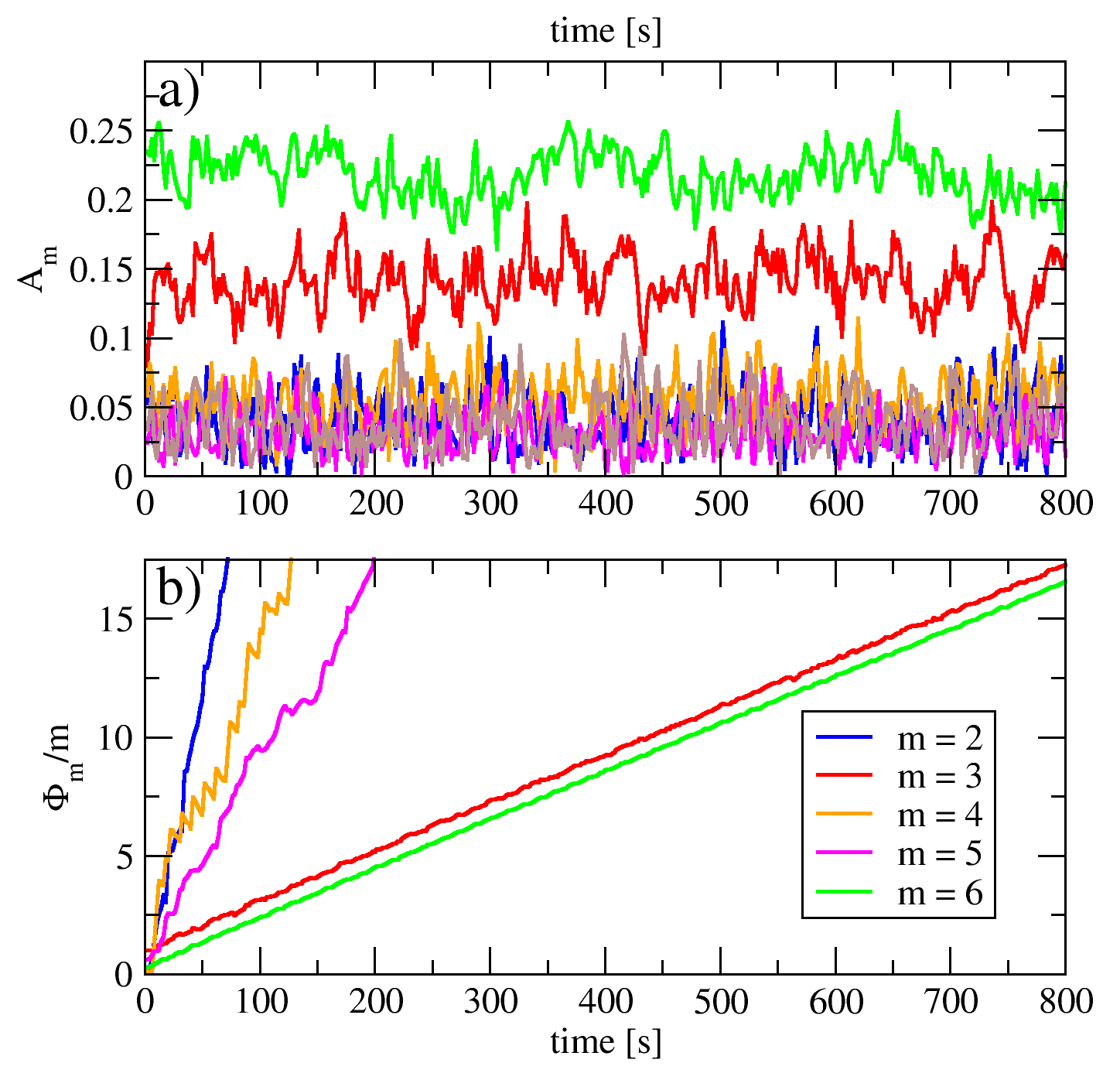}
 \centering
\caption{Temporal development of the Fourier amplitudes (a) and ``azimuthal distances'' (b) of wave modes $m=2,\dots,6$ in a laboratory experiment ($\Omega=4.2$ rpm, spin-down series). Note, that the modes of the dominant wave number $m=3$ and its ``slave pattern'' $m=6$ -- that has the largest amplitude -- exhibit regular, uniform drift, whereas the small-amplitude modes provide bogus `non-physical' signals in the bottom panel.}
\label{prop_demo}
\end{figure}

Next, the drift rates of the dominant wave modes were determined and analyzed.
The discrete Fourier transform, described in the ``Methods'' section, yielded the phase shifts $\phi_m$  for each time instant.
Thus, the quantity $\phi_m(t)/m$ could be used as a measure of the ``azimuthal distance'' travelled by the given component with wave number $m$ since $t=0$. Such time series are shown for the two largest Fourier components ($m=3$ and $m=6$) in the explanatory figure Fig. \ref{prop_demo}b obtained in a laboratory experiment ($\Omega=4.2$ rpm, spin-down series), alongside with amplitudes $A_m(t)$ of the first six Fourier components in Fig. \ref{prop_demo}a. For better visualization of the evolution of $\phi_m(t)/m$ in the bottom panel, we extended the periodical $[0;2\pi]$ range to $[0;+\infty)$ (so that the positive increments correspond to counter-clockwise propagation).

The drift rate $c_m(t)$ (angular velocity) of a given mode $m$ could thus be obtained as the slope of the corresponding graph at time instant $t$, since:
\begin{equation}
\frac{1}{m}\frac{\partial \phi_m(t)}{\partial t}\equiv c_m(t),
\label{omega}
\end{equation}    
therefore linear fits to the quasi-stationary part of the propagation could be used to determine $c_m(t)$.

It is important to mention, that  within a given experiment all the Fourier components of significant amplitudes propagated at the same drift rate, i.e. no wave dispersion was present. Consequently, although the flow pattern drifted around the annulus, its form remained unchanged. 
We note, that in a previous experimental series carried out in the same set-up with the addition of \emph{sloping bottom topography}, marked wave dispersion was observed. In that case, the stable baroclinic wave patterns emerged in the form of so-called resonant triads \citep{our_npg}. Moreover, \cite{pfeffer} also found dispersion in their flat bottom experiment, and \cite{uwe_piv} reported dispersion in the wave transition region of the $Ta-Ro_T$ diagram at lower $\Delta T$.

\begin{figure*}
\noindent\includegraphics[width=14cm]{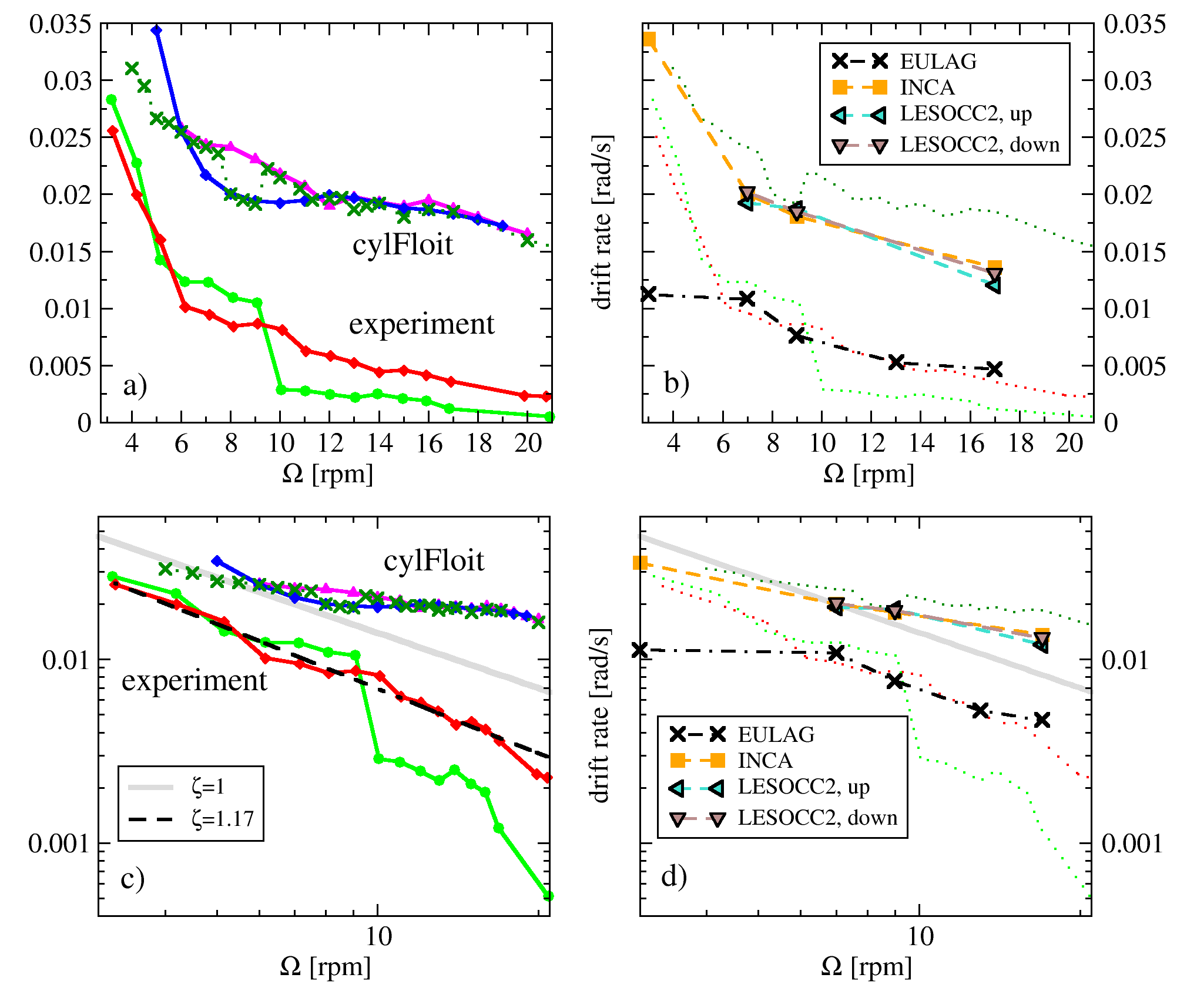}
 \centering
\caption{Drift rates of the dominant wave modes as functions of rotation rate $\Omega$. In panel (a), the experimental spin-up (green), spin-down (red) sequences are presented, alongside the spin-up (magenta), spin-down (blue) and ``from scratch'' (dark green) series. In panel (b) the drift rates from other numerical models are shown. For a better comparison, three curves of panel (a) are repeated here with dotted lines, using their original color coding. The data from panels a) and b) are repeated with double logarithmic scales in panels c) and d). The power-law fit of the (spin-down) experimental data points (dashed line) and $\zeta=1$ curves (grey) obtained via thermal wind balance are also shown.}
\label{prop_results}
\end{figure*}

We compared the drift rates of the wave mode of the largest average amplitude $\langle A_m(t) \rangle$ for each run. 
The drift rates obtained for the laboratory experiments are presented in Fig. \ref{prop_results}a, both for the spin-up (green) and spin-down series, as a function of rotation rate $\Omega$. An overall decreasing trend can be observed in agreement with the expectations based on quasi-geostrophic theory.
Due to thermal wind balance, the velocity of the zonal background flow is expected to scale as:
\begin{equation}
U\propto\frac{\alpha g D \Delta T}{2\Omega (b-a)}.
\end{equation} 
In the linear theory of \cite{eady} the baroclinic waves themselves also propagate at the velocity of the mean flow, thus a $c_m\propto \Omega^{-1}$ scaling is to be expected. Accordingly, \cite{fein} found in baroclinic annulus experiments the general power-law form $c_m=B(\alpha\Delta T/\Omega)^\zeta$.
In the case of our experiments (the spin-down series was evaluated), these parameters were found to be $B=4.4\pm 0.15$ and $\zeta=1.17\pm 0.04$. The fit is shown in Fig. \ref{prop_results}c -- the repetition of panel a) with logarithmic scales -- as a dashed line, and a $\zeta=1$ slope corresponding to the thermal wind speed is also plotted (thick grey curve). It is to be noted, that for a free-surface annulus Fein obtained $\zeta=0.88\pm0.07$ (the values of $B$ are not suitable for direct comparison between different set-ups as they depend on the actual geometrical parameters of the tanks used).  

Fein also demonstrated that both in terms of factor $B$ and exponent $\zeta$ the experiments with free surface and rigid lid exhibit significantly different scaling properties, leading to an order-of-magnitude difference between their respective drift rates (the waves in the free surface set-up being the faster). This observation underlines the extreme sensitivity of the studied system to the upper boundary condition, and thus gives a broader context to our comparisons with the numerical results, which now follows.  

The $c_m$ values, obtained from the cylFloit data are shown with magenta and blue curves in Fig. \ref{prop_results}a and c, representing the drift rates in the spin-up and spin-down series, respectively. 
Also, the results from the ``from scratch'' series (always initiated from the stable $m=0$ state) are plotted with a blue graph.
Figure \ref{prop_results}b and d show the drift rates found in the LESOCC2 (spin-up and spin-down), INCA and EULAG simulations.
The general decreasing trend of drift rates were captured by the investigated models, and the drift rates of cylFloit, INCA and LESOCC2 are in fairly good agreement with each other, yet, neither the experimentally obtained, nor the thermal wind-type scaling was reproduced by them. The drift rates are generally overestimated compared to the laboratory findings (the experimental curves and the cylFloit ``from scratch'' points are repeated in \ref{prop_results}b and d in the form of a dotted curves, and a $\zeta=1$ power-law is also given in panel d). The EULAG simulations however -- aside for the $\Omega=3$ rpm case, where the wave pattern appeared rather irregular -- were in good agreement with the experiments in terms of drift rates. The possible reasons for these differences will be discussed in the ``Summary and conclusions'' section.

Besides the general decreasing trend of $c_m$, the most marked feature in the experimental spin-up sequence (green curves in \ref{prop_results}a and c) is a sharp drop around $\Omega=10$ rpm, a data point which lies well within the regime of dominant wave number $m=3$ (cf. Fig. \ref{subway}a).
This transition was also observed in terms of the average thermal variance $\bar{\sigma}$, as mentioned in the previous subsection.

It is to be noted, that in the experimentation procedure, the discussed drop coincided with an interruption of the measurement sequence. The spin-up measurements were conducted in four campaigns on subsequent days.
The measurement protocol in such cases was the following: on a new measurement day, the spin-up process was repeated from an initial axially symmetric state with a fully established sideways convection ($\Omega \approx 2$ rpm), up to the preceding data point (in this case to $\Omega=9.1$ rpm), which was then left undisturbed for a long relaxation time (here 4 hours and 40 minutes). Afterwards, the standard spin-up procedure -- described in the ``Methods'' section -- was conducted to approach the new parameter point (in this case: $\Omega = 10.1$ rpm).
Interestingly, this was the single case where the re-initiation of the measurement sequence coincided with such an abrupt change. Similar interruptions and re-initiations occurred between the data points of $\Omega=4$ rpm and $\Omega=5$ rpm and between $\Omega=15$ rpm and $\Omega=16$ rpm (and also, between $\Omega = 15$ rpm and $\Omega = 14$ rpm in the spin-down series), without any significant effect on the drift rates.

\begin{figure}
\noindent\includegraphics[width=7cm]{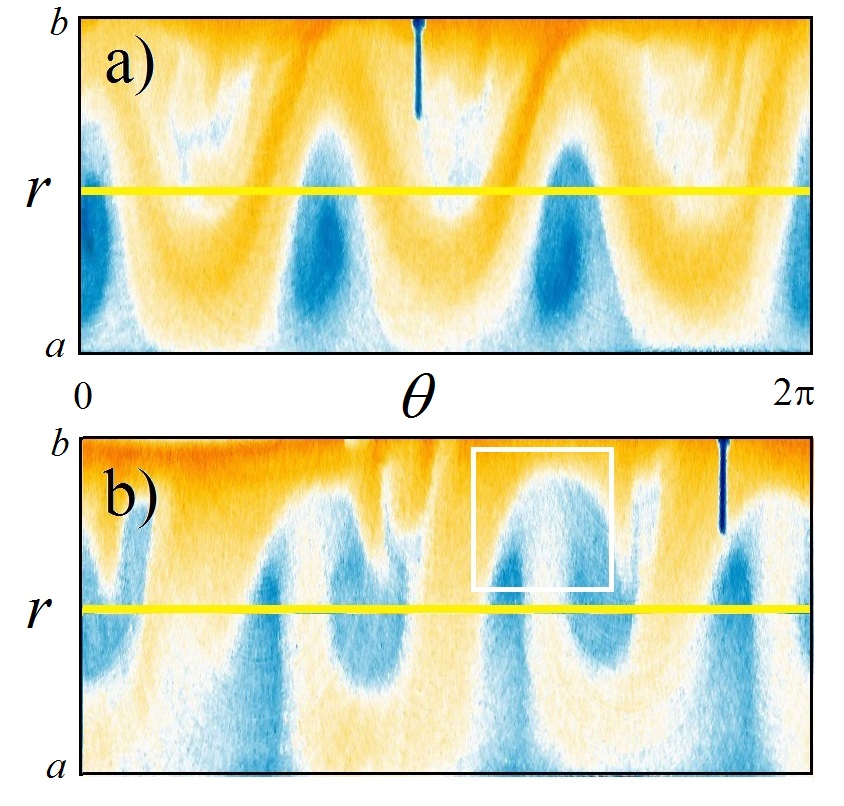}
 \centering
\caption{Two thermographic experimental snapshots of $m=3$ surface temperature patterns. A fastly propagating type (a), observed at rotation rate $\Omega=4.2$ rpm (see also the corresponding propagation plot in Fig. \ref{prop_demo}b),
and the slower type, observed after the ``topological transition'' ($\Omega = 10.1$ rpm).}
\label{two_states}
\end{figure}

As mentioned above, the observed phenomenon was not accompanied with the change of the dominant wave number, yet, a certain  \emph{topological transition} of the surface temperature field was detected.
Fig. \ref{two_states} shows two typical snapshots, transformed to polar coordinates. The pattern characteristic to the first, ``classic'' type of $m=3$ waves (observed in the range of 7.1 rpm $\leq \Omega \leq$ 9.1 rpm) is presented in panel a), whereas the structure of the slowly propagating type (10.1 rpm $\leq \Omega \leq$ 15.9 rpm) is visible in panel b).     
One can observe, that the neighboring cold eddies that are separated by the meandering warm jet in case a), are connected by cold filaments in case b) (e.g. the one in the white rectangle). This implies that the widely used experimental classification of baroclinic waves in a rotating annulus -- that is based on the dominant wave number only -- is rather incomplete, since apparently even if $\Omega$, $\Delta T$ and $m$ are given, clearly different dynamical states may develop that essentially have the \emph{same} dominant zonal wave number.     
 
Similarly to the experimental data, a pronounced hysteresis appears at rotation rates $\Omega < 13$ rpm in the cylFloit results (Fig. \ref{prop_results}a).
In this case the $\Omega$-range of the hysteretic regime clearly agrees with the one found in terms of the dominant wave numbers (cf. Fig. \ref{subway}b). The interval between the intersection points of the spin-up and spin-down curves ($\Omega=6$ rpm and $\Omega=12$ rpm) can therefore be described as the regime where  $m=4$ is the dominant mode of the (lower) spin-down branch \emph{and} the (upper) spin-down branch exhibits $m=3$. Thus, a manifest correlation is present: at a given $\Omega$ the waves of three-fold symmetry propagate faster than the four-fold-symmetric patterns.
This conclusion is confirmed by the behavior observed in the from-scratch-initiated simulations of the dark green curve (see also the blue curve of Fig. \ref{subway}b): in the hysteretic regime, when the system switches from one branch to the other in terms of $m$, it does so in the drift rate as well.
Note, that below $\Omega=10.1$ rpm (where the aforementioned topological re-organization and sudden drop in the drift rates took place), also in the experimental data of Fig. \ref{prop_results}a, the intersection point of the two branches coincides with the onset of the $m=3$ mode in the spin-up sequence, whereas the spin-down branch maintains the dominant wave number of $m=4$. In other words: the ``first'' type of $m=3$ patterns (seen in Fig. \ref{two_states}a) drifts faster than the baroclinic waves of $m=4$ at a given rotation rate $\Omega$.

\subsection{Empirical Orthogonal Functions}
To properly describe the temperature variance stored in co-existent spatio-temporal patterns in the annulus we turned to the method of Empirical Orthogonal Functions (EOFs) \citep{uwe_agu}. This approach is generally accepted as a powerful tool for data compression and dimensionality reduction: it is able to find the spatial patterns of variability, their time variation, and provides a measure for the ``relevance'' of each pattern, and thus describe the complex behavior of the system, often in terms of surprisingly few modes \citep{eof_book}. It is to be noted, however, that in general these EOF modes do not necessarily correspond to individual dynamical eigenmodes of the system \citep{eof_review}.

\begin{figure*}
\noindent\includegraphics[width=14cm]{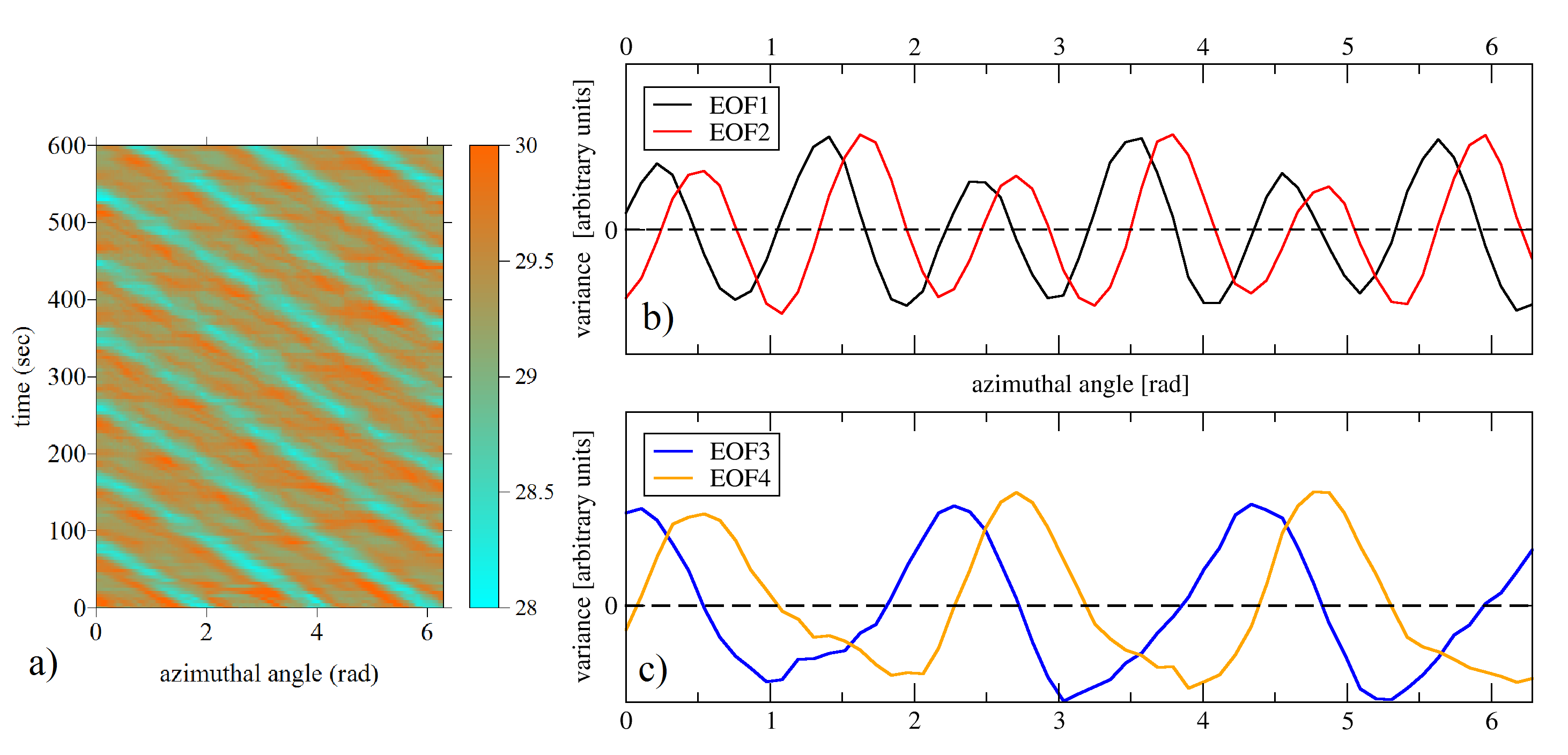}
 \centering
\caption{A typical thermografic Hovm\"oller (space-time) plot of an experimental run at dominant wavenumber $m=3$ (a), and the first two corresponding EOF variance pattern pairs (b and c). The corresponding relative variances of EOFs 1 to 4 were $p_1=0.29$, $p_2=0.27$, $p_3=0.082$ and $p_4=0.073$, respectively.}
\label{eof_demo}
\end{figure*}

EOF analysis has been extensively used in recent works \citep{uwe_piv, borchert} for two-dimensional temperature and velocity fields in the particular setup at BTU CS. Here, however, as we restricted our studies to the temperature profiles along the circular contour at mid-radius, the one-dimensional EOFs were determined.
Organizing the surface temperature data $T(\theta, t)$ at given time instants as column vectors (state vectors) and combining them in temporal order, yields the so-called data matrix $\mathbf{X}$, whose number of rows and columns correspond to that of the considered spatial and temporal points, respectively. In the present one-dimensional case a transparent visual representation of $\mathbf{X}^T$ can be obtained in the form of a space-time or Hovm\"oller plot, e.g. the one shown in Fig. \ref{eof_demo}a (corresponding to an $m=3$ baroclinic wave).

In our EOF analyses the selected matrices $\mathbf{X}$ consisted of the data from the last 100 time instants of the given (either experimental or numerical) run; a time interval that always lied well within the quasi-stationary part of the investigated process. In space, the experimental data were linearly interpolated onto an azimuthally equidistant grid of 100 cells, whereas the numerical data were transformed similarly to 50 grid points of uniform spacing.  
The entries of $\mathbf{X}$ were then obtained by subtracting the mean value of each corresponding row (i.e. temperature time series at a given spatial location).      
The covariance matrix $\mathbf{S}$ is given by:
\begin{equation}
\mathbf{S}=\frac{1}{n-1}\mathbf{X}\mathbf{X}^T,
\label{cov_mtx}
\end{equation} 
where $n=100$ is the number of time instants considered.
The eigenvectors $\mathbf{e}_k$ (i.e. the EOFs themselves) and the corresponding eigenvalues $\xi_k$ of $\mathbf{S}$ were computed. The EOF index $k=1,2,3,\dots$ is given by organizing the eigenvalues in decreasing order as: $\xi_1 \geq \xi_2 \geq \xi_3\geq \dots$. The percentage contribution $p_k$ of each pattern $\mathbf{e}_k$ to the total variance captured by the EOFs can then be expressed as: $p_k = \xi_k / \sum_i \xi_i$. As a demonstration, the first four EOF patterns are shown in Fig. \ref{eof_demo}b, corresponding to the same experiment as the Hovm\"oller plot of panel a).
     
\subsubsection{Variance distribution}
\begin{figure*}
\noindent\includegraphics[width=16cm]{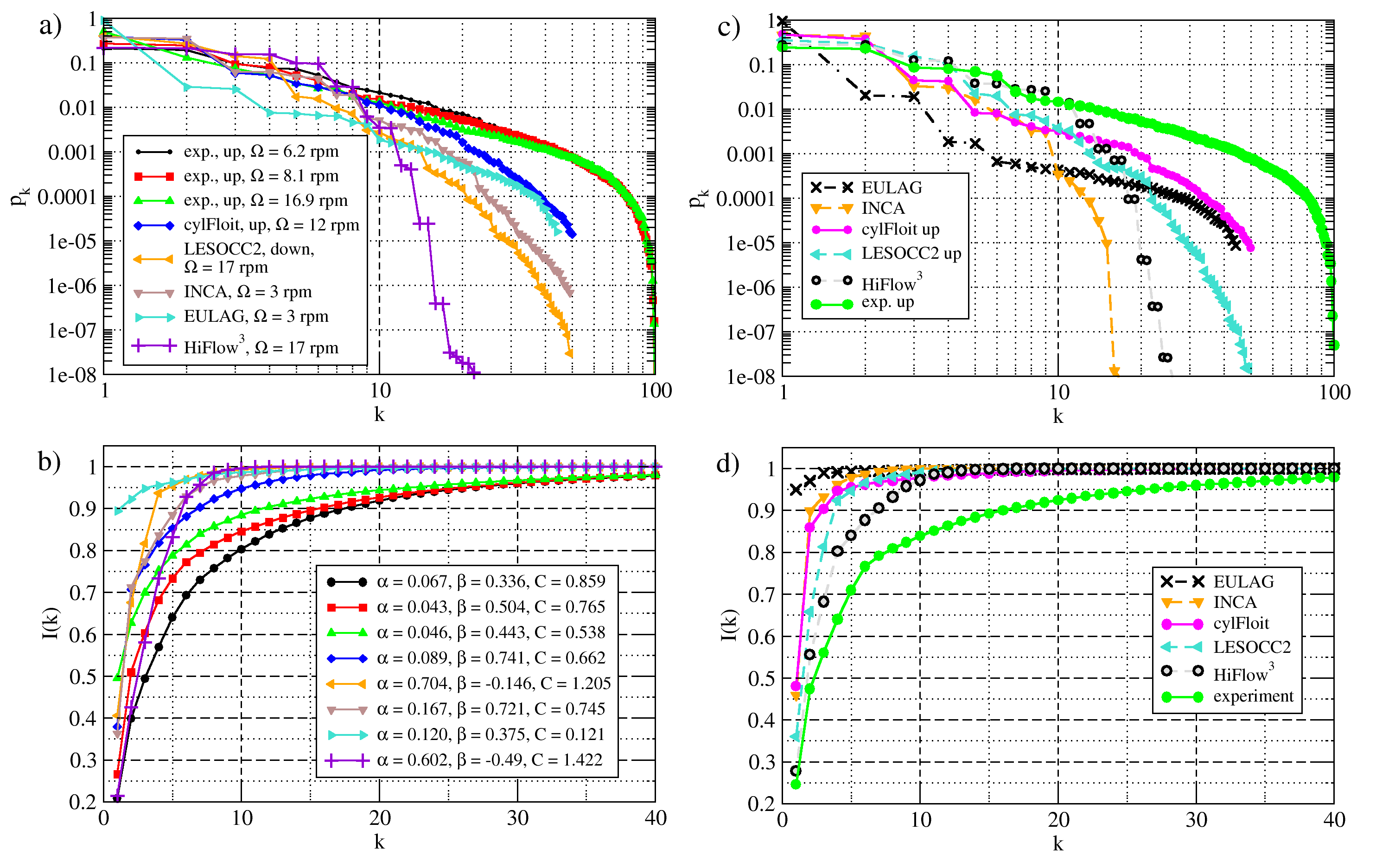}
 \centering
\caption{Typical variability density functions obtained from the experiments and numerical models (a). (See legends for the model types and rotation rates). Their corresponding cumulative density functions are shown in panel (b) with the same color coding. The fitted parameter values of $\alpha$, $\beta$ and $C$ are also shown.
Panels c) and d) show the density functions and cumulative density functions, respectively, for all the models and the experiment for the $\Omega=9$ rpm (spin-up) case.}
\label{pdf_cdf}
\end{figure*}

The distribution of percentage contributions $p_k$ of the EOFs (a monotonically decreasing function of index $k$) was analyzed to quantify the overall complexity of the investigated spatio-temporal patterns.
Typical ``variability density functions'' are presented in Fig. \ref{pdf_cdf}a, as obtained from our experiments (black, red and green curves) and the simulations with different models (see also the legend).
It is to be emphasized that this figure serves a purely explanatory purpose: to help the reader to better understand the role of the parameters used to quantify the distribution properties. Therefore a large variety of cases at different rotation rates are shown, which are therefore not meant for model comparison.
Yet, some common features can be observed: visibly, in the most of the domain, experimental data points exhibit a power-law type scaling -- indicating the importance of higher EOF indices -- that is followed by exponential cut-off.
A qualitatively similar behavior can be observed in the numerical data as well, however, both the ``power-law part'' and the ``cut-off part'' appear to have different quantitative properties than the ones of the experimental results.

To find appropriate measures of these properties, firstly the cumulative density functions $I(k)=\sum_{i=1}^{k}p_k$ were calculated for each experimental and numerical run. Fig. \ref{pdf_cdf}b shows the $I(k)$ curves corresponding to the cases plotted in panel a), with the same color coding.
The heuristic empirical form 
\begin{equation}
I(k)=1-C\frac{e^{-\alpha\cdot k}}{k^\beta}
\label{cdf}
\end{equation} 
has proven to be a strikingly accurate parametrization for every run: typically, the asymptotic standard errors were below $3\%$ for all three free parameters $\alpha$, $\beta$ and $C$. 
Note, that the values of these parameters for the exemplary cases of Fig. \ref{pdf_cdf}b are listed in the legend. 
In panels c and d the density functions and cumulative density functions of all the models (and the experiment) are given, all for a single parameter point $\Omega\approx 9$ rpm. For all models, the values of $\alpha$, $\beta$ and $C$ were evaluated for each simulated $\Omega$.

\begin{figure}
\noindent\includegraphics[width=8cm]{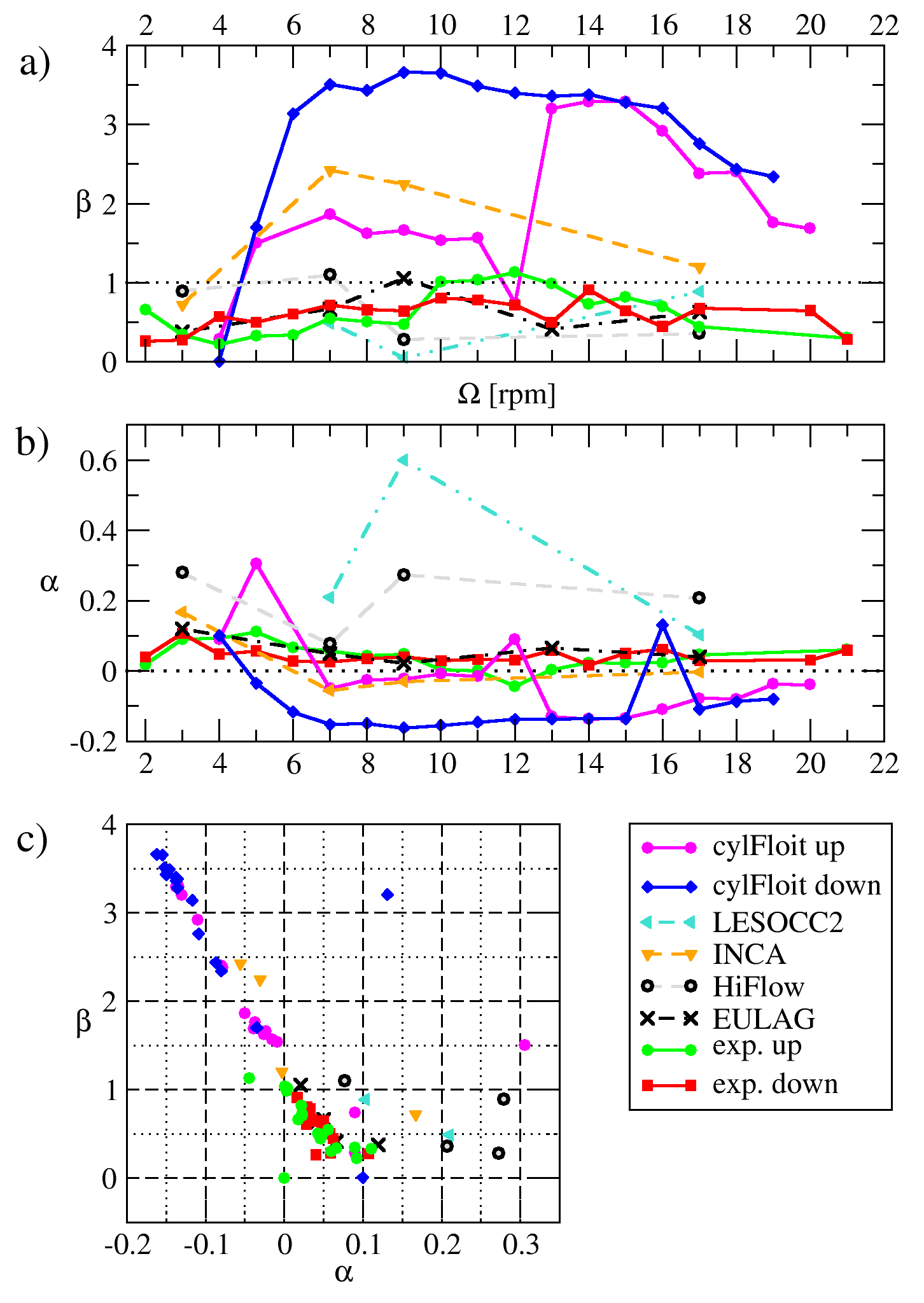}
 \centering
\caption{The fitted parameters $\beta$ and $\alpha$ of the cumulative density functions as functions of rotation rate $\Omega$: panels a) and b), respectively, and the correlation plot of the two parameters (c). The color coding is the same for all panels.}
\label{alpha_beta}
\end{figure}

Let us now compare the fitted parameters $\beta$ and $\alpha$ versus rotation rate $\Omega$ in Figs. \ref{alpha_beta}a and b, respectively. In the laboratory experiments (red and green curves in both panels) the values of $\beta$ scatter in the range  of $\beta \in (0.3;1.1)$, while $\alpha$ exhibits small positive values $\alpha \in (0.01;0.1)$. These imply that the saturation of the cumulative density function is slow, a considerable part of the variance is stored in the EOFs of larger $k$. As the exponential factor is such a slowly varying function (due to the small $\alpha$), the behavior observed in the experimental density functions of Fig. \ref{pdf_cdf}a approximately follows a power-law scaling in the form of $ k^{-\gamma}\equiv k^{-\beta-1}$ with $1.3<\gamma <2.1$. Such values of $\gamma$ are typical for the probability density functions of long-range correlated processes. As yet another measure of complexity, it is to be mentioned that  $k=6-18$ different EOFs were needed to cover $90\%$ ($I(k)=0.9$) of the total variance in the experimental distributions (like the first three graphs listed in Fig. \ref{pdf_cdf}b).

The exponent $\beta$ was also typically found within the same $0 < \beta < 1$ regime in the simulations conducted by EULAG, HiFlow$^3$ and LESOCC2 (see the black, gray and turquoise graphs in Figs.\ref{alpha_beta}a, respectively). This implies that the distribution of variance in these three models behave realistically concerning the smaller $k$-regime, which practically corresponds to the large-scale features of the flow. Also in terms of $\alpha$, the EULAG results scattered perfectly within the same interval as the experiments, meaning that the ``tail'' of the distribution scales correctly.
However, the values of parameter $C$ were an order of magnitude smaller for EULAG ($C\in(0.025;0.12)$) than for all the other cases, either experimental or numerical, where $C\in(0.42;1.17)$ within the baroclinic unstable regime. This is due to the interesting fact that in these simulations -- despite of their close-to-perfect scaling properties -- the very first EOF alone was responsible for $90-96\%$ of the total variance, i.e. $p_1 \in (0.9;0.96)$, a property that can be observed on the turquoise curve of Fig. \ref{pdf_cdf}b too. For HiFlow$^3$ and LESOCC2, on the other hand, parameter $\alpha$ appeared to be 2-6 times larger than in the experiments (Fig. \ref{alpha_beta}b), meaning that the variability of larger indices $k$ is suppressed by a marked exponential cut-off, thus most of the variance is stored in the large-scale patterns.

The INCA and cylFloit model runs generally exhibited significantly larger values of exponent $\beta$ than the laboratory results (see orange, blue and magenta graphs in Fig. \ref{alpha_beta}a). Typically, the cases where $\beta > 1$ holds, correspond to $\alpha < 0$, as visualized in the correlation plot of Fig. \ref{alpha_beta}c. This relation suggests that at smaller values of index $k$ a sharp ``fast'' power law characterizes the dominant, large-scale part of the distribution. This scaling, however, is confined only to this regime: in itself it would mean a too sharp cut-off at larger indices $k$. Thus, for an appropriate parametrization, a negative value of $\alpha$ is needed to compensate this effect to keep the variances at higher EOF indices finite. 

Regarding the cylFloit simulations the data points of the spin-up and spin-down series are plotted separately, with magenta and blue symbols in all panels of Fig. \ref{alpha_beta}, respectively. In panel a) the marked hysteretic behavior of parameter $\beta$ can be observed. This behavior is in manifest connection with the dominant wave numbers (cf. Fig. \ref{subway}b): apparently, $m=4$ states are characterized by larger $\beta$ than $m=3$ states. This implies that in the $m=4$-dominated states the ``scale separation'' is more pronounced: a larger fraction of the total variance is stored in the first few EOF modes than in the $m=3$ cases. 

In the spin-up and spin-down sequences of the laboratory experiments no such connection was found between wave numbers and the parameters of $I(k)$, however, a significant jump of $\beta$ at rotation rate $\Omega=10.1$ rpm is visible in the spin-up curve (green graph in Fig. \ref{alpha_beta}a), that corresponds to the topological transition within the $m=3$ regime, described in the previous section.     
 
\subsubsection{Pattern correlations}
Besides the distributions of the eigenvalues of covariance matrix $\mathbf{S}$, the eigenvectors $\mathbf{e}_k$, i.e. the variance patterns themselves were also compared. The applied method was similar to the one used in \cite{borchert} for two-dimensional EOFs. Firstly, the obtained EOF patterns of indices $k$ and $l$ from the experiment and a given numerical model were linearly interpolated onto the same equidistant grid of 100 cells. These functions are marked by: $f_k^{\rm exp}(\theta)$ and $f_l^{\rm mod}(\theta)$, respectively ($\theta\in(0;2\pi]$).
Their correlation coefficient is then calculated as:
\begin{equation}
C_{kl}=\frac{\langle f_k^{\rm exp}(\theta) f_l^{\rm mod}(\theta+\varphi)\rangle-\langle f_k^{\rm exp}(\theta)\rangle \langle f_l^{\rm mod}(\theta+\varphi) \rangle}{\sigma(f_k^{\rm exp}(\theta))\sigma(f_l^{\rm mod}(\theta+\varphi))},
\end{equation} 
where $\langle\cdot\rangle$ marks the azimuthal mean, $\sigma(\cdot)$ denotes the standard deviation and $\varphi$ is the ``offset angle'' which maximizes $C_{kl}$. This sliding transformation is required due to the fact that the azimuthal orientation of EOFs in the various models (and experimental runs) are generally different. In this transformation periodic boundary conditions were applied, i.e. the values for which $\theta+\varphi>2\pi$ were actually mapped onto the interval $(0;\varphi)$. 

\begin{figure*}
\noindent\includegraphics[width=16cm]{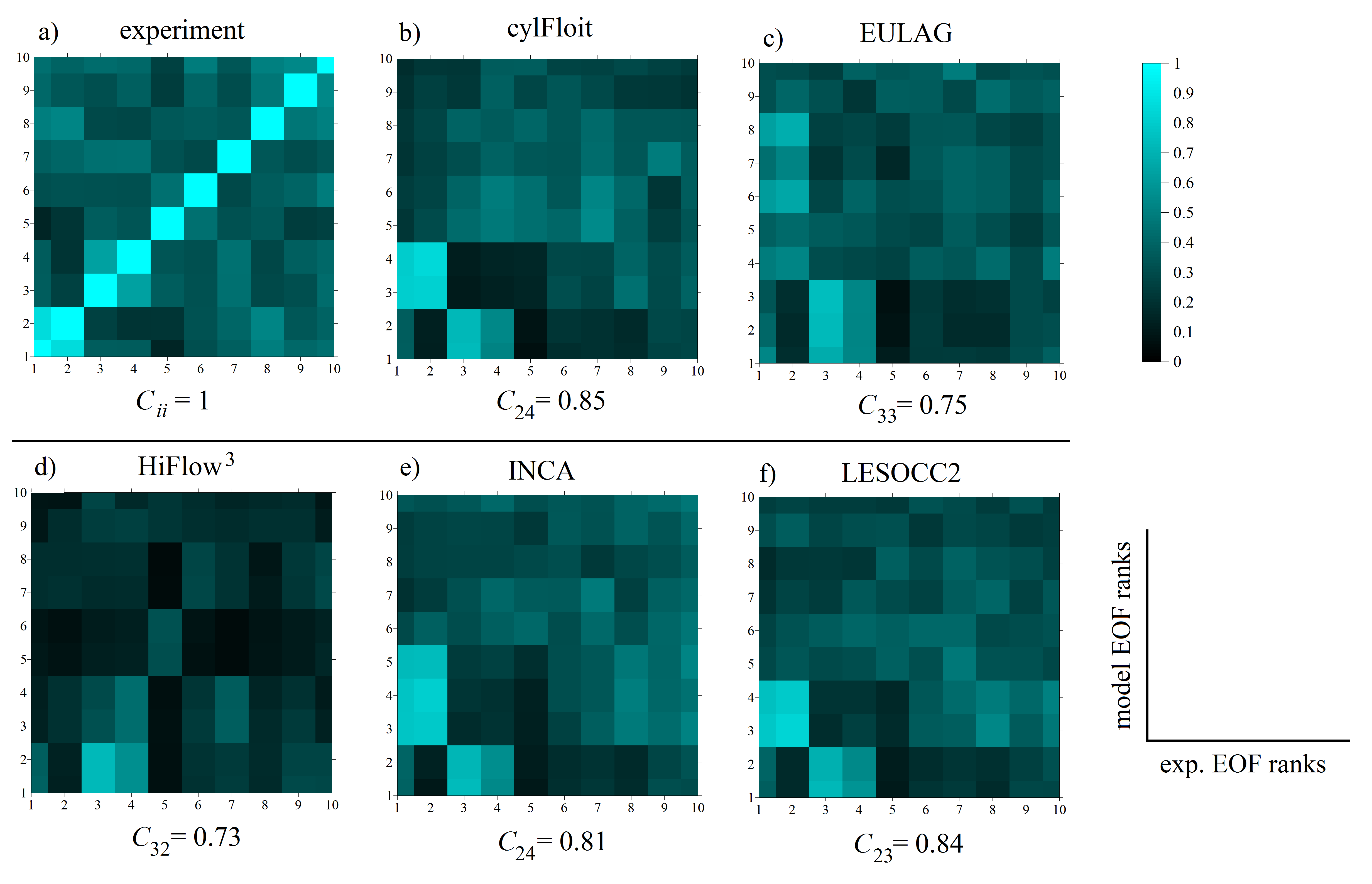}
 \centering
\caption{The cross-correlation matrices obtained in the benchmark case $\#4$ ($\Omega \approx 17$ rpm). The positions and values of the maximum entries of the matrices are also given underneath the respective figures.}
\label{corr_matrices}
\end{figure*}

The values $C_{kl}$ were calculated for the first 10 EOFs (both numerical and experimental) and were combined into $10\times 10$ matrices. The structures of these matrices were analyzed. Here, we present a few typical exemplary cases to yield a qualitative insight to the nature of the correlation properties of one-dimensional EOFs. 
In Fig. \ref{corr_matrices} the correlation plots for the benchmark case $\#4$ ($\Omega \approx 17$ rpm) are presented. This case was selected due to the fact that here -- already out of the hysteretic regime -- all of the models found $m=4$ as dominant mode, in agreement with the experiments.   
For a better understanding of the comparisons to follow, in panel a) we present the correlation plot of the EOFs of the given experiment with one another (hence, $f_i^{\rm exp}\equiv f_i^{\rm mod}$ using the above notation). Trivially, in this case $C_{ii}=1$ holds for the diagonal entries, and the matrix is symmetric. Though the EOFs are, by definition, orthogonal, yet, the aforementioned sliding transformation leads to rather marked correlations, due to the fact, that the EOF1 and EOF2 (and, similarly EOF3 and EOF4, etc.) are rather similar, but shifted in azimuthal direction (see also Fig. \ref{eof_demo}b and c). Such EOF pairs account for the baroclinic wave propagation, analogously to the relation of sine and cosine terms in the Fourier decomposition of propagating waves.

Panels b)-f) show the correlation matrices obtained from the comparison of the experimental set of EOFs with the EOFs from cylFloit, EULAG, HiFlow$^3$, INCA and LESOCC2, respectively. The numbers on the horizontal axis represent the indices of the experimental variance patterns, and those on the vertical axis are the EOF indices of the given numerical model.   
The indices and values of the maximum entries in the given matrix are also marked in the panels.
Two main observations need to be emphasized. Firstly, the structures of the matrices are rather similar, implying that the numerical models produce similar variance patterns to one another. Also, the aforementioned EOF pairs are clearly visible in the matrices in the form of $2\times 1$ and $2\times 2$ blocks of closely similar correlations.
The second main observation is that, despite of the similarity of the matrices, none of them has diagonal structure. Thus, the various EOF patterns are ranked differently. 

\begin{figure*}
\noindent\includegraphics[width=14cm]{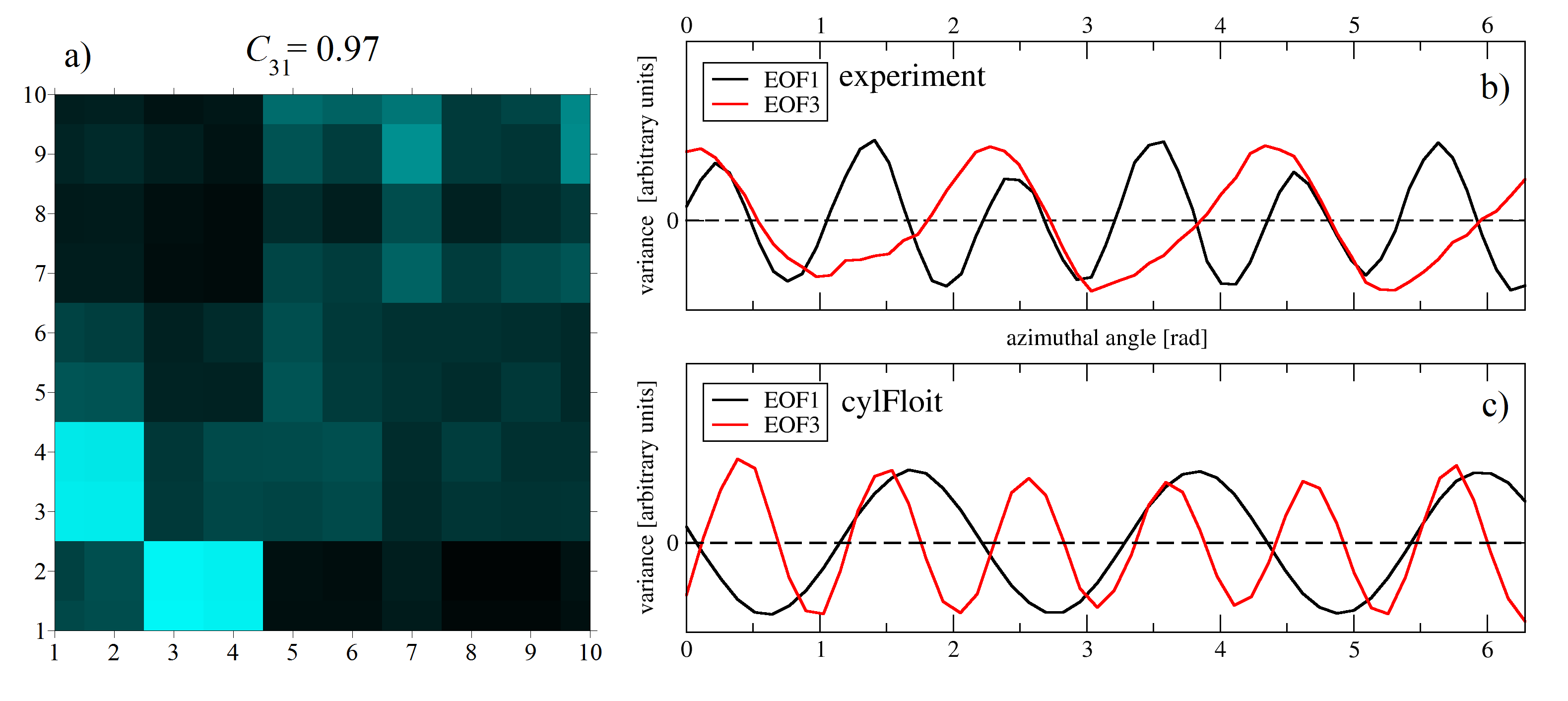}
 \centering
\caption{The correlation matrix of the one-dimensional EOFs, obtained from the numerical and experimental data of test case \#7 of \cite{borchert}, and the value of the maximum entry (a). EOFs 1 and 3 of the experimental (a) and numerical (b) case.
Note the ``swap'' between the indices and wave patterns of the two cases.}
\label{historic_test}
\end{figure*}

The latter statement is seemingly in contrast with the findings of \cite{borchert}, who found correlation coeffitiens above $0.9$ by comparing their EOFs (obtained using the cylFloit and INCA codes) to the laboratory EOFs of the \emph{same} index. However, there the full two-dimensional surface temperature patterns were taken. As a test of consistency, we applied our methodology to the very same experimental records from year 2011 and the same (``from-scratch'' initiated) cylFloit runs studied in \cite{borchert} to obtain the correlation coefficients for the one-dimensional EOFs. The resulting correlation matrix is shown in Fig. \ref{historic_test}a. Apparently, the obtained structure is quite similar to those seen in Fig. \ref{corr_matrices}b-f, and lacks large values in the diagonal. However, the entries in the $2 \times 2$ blocks in he vicinity of the diagonal at lower left are indeed large, with a maximum of $C_{31}=0.97$. The similarities and differences of these patterns can be visually evaluated in Fig. \ref{historic_test}b and c, where EOFs 1 and 3 are plotted for the experimental and the numerical case, respectively. One can see, that in the experimental case EOF1 exhibits wave number $m=6$ (and so does its shifted pair EOF2, not shown here) and the dominant baroclinic wave number $m=3$ appears in the EOF3 for the first time, in contrast to the typical numerical results. Thus, the numerical models have a tendency to underestimate the variance stored in the smaller scales.

It can be stated that the one-dimensional data extracted from the surface temperature field at mid-radius $r_{\rm mid}$ are generally more sensitive to smaller-scale differences than the full two-dimensional patterns, since -- as discussed above -- in the two-dimensional case no such ``EOF swap'' occurs between numerics and experiment. The mid-radius temperature profiles are apparently largely effected by the variance stored in the harmonics of the dominant baroclinic wave mode, related to the structure and dynamics of the cold eddies in the lobes of baroclinic waves. The fact that the numerical models are apparently not able to resolve these phenomena implies that they may well be related to boundary layer effects or even ``wind'' stress above the free surface of the laboratory tank, which are clearly out of the scope of the investigated numerical models.        

Also, it is to be noted, that in an annulus with an exact rotational invariance the EOFs must be sinusoidal, i.e. each would project on a single azimuthal wave number only, as shown by \cite{achatz}. The fact that the typical EOFs of the experiment can in many cases visibly be decomposed to at least two wave numbers (as the ones in Fig. \ref{eof_demo}b and Fig. \ref{historic_test}b) indicates a violation of rotational symmetry and nonlinear dynamics. In the azimuthally invariant numerical models (as cylFloit), however, the EOF patterns were indeed found to be nearly sinusoidal (see e.g. Fig. \ref{historic_test}c). Their slight imperfections is merely a consequence of the finite length of the time series considered.

\section{Summary and conclusions}
In this work we have critically compared various experimentally and numerically obtained characteristic properties of baroclinic instability in a differentially heated rotating annulus. Our systematic comparison of five different numerical models to laboratory experiments (``benchmarking'') was largely motivated by the general need to validate numerical models and procedures to be used for modeling large-scale atmospheric flows. 

Two series of laboratory measurements were performed: the ``spin-up'' and ``spin-down'' sequences. Between each measurement only rotation rate $\Omega$ was adjusted, while the radial temperature difference $\Delta T\approx 8$ K remained constant. The two sequences enabled us to scan through the investigated parameter range with different initial conditions, and thus access multiple equilibrium regimes. In agreement with earlier results \citep{miller_butler,christoph_book,thomas_npg} a considerable hysteresis was found in terms of the dominant azimuthal wave numbers $m$ of the baroclinic waves.   

It is well established since the works of \cite{james} and \cite{hignett} in the 1980s, that in terms of $m$, the development of baroclinic waves in baroclinic annuli can be captured in direct numerical simulations fairly well. In the present work we also found that $m$ is indeed a robust indicator of the flow state, and its obtained values exhibit good agreement between the experiments and the numerical runs. The numerical results also support our conclusion that the hysteretic behavior of $m$ is to be interpreted as distinct multiple equilibria (bifurcation) and is not just caused by transient phenomena. This statement is backed by the following observations: (i) Simulation series conducted with models cylFloit and LESOCC2 imitated the ``spin-up'' and ``spin-down'' sequences and found hysteresis in terms of $m$. A third bunch of simulations, however, were always initialized from the axially symmetric stable state (cylFloit ``from scratch'' sequence). Yet, occasionally even here, wave numbers characteristic to the ``spin-down'' branch were found to develop within a rotation rate regime where these simulations typically converged to the states of the ``spin-up'' branch. (ii) In the HiFlow$^3$ simulations, runs with slightly perturbed initial conditions were also conducted. The only cases where these temperature disturbances yielded different dominant wave number $m$ than the corresponding unperturbed runs were at parameter points within the experimentally observed hysteretic $\Omega$-regime.                 

Another important measure of baroclinic wave dynamics is the drift rate $c_m$ of the dominant wave mode. In qualitative agreement with the quasigeostrophic Eady model \citep{realvallis}, the $c_m(\Omega)$ relationship was found to be a decreasing function, roughly following the $c_m \propto \Omega^{-1}$ dependence set by the thermal wind balance. It is to be noted, however, that most of the models (with the exception of EULAG) systematically over-estimated the wave speeds. This phenomenon may well be explained by the simulations difficulties to resolve the boundary layer drag at the lateral sidewalls. A similar observation was described in the study of \cite{williams} where a two-layer (lid shear-driven) rotating baroclinic annulus set-up was investigated both experimentally and numerically. In their case the simulated drift rates were larger than the measured values by a factor of 4, due to the model's neglect of Stewartson layer drag. Stewartson layers are characteristic for homogeneous fluids. In our case of relatively strong stratification, however, $Pr\,Ro_T/\Gamma^2\gg Ek^{2/3}$ holds with $\Gamma$ being the vertical aspect ratio of the tank (as defined in Section 2) and $Ek=\nu/(\Omega L^2)$ the Ekman number. In this regime -- instead of Stewartson layers -- two boundary layers are found in the vicinity of each lateral sidewall: the larger hydrostatic layer with a characteristic thickness of $\delta_h=D(Pr\,Ro_T/\Gamma^2)^{1/2}$ and, closer to the wall, the buoyancy layer whose thickness is $\delta_b=D(\nu\kappa/(D^3g\alpha\Delta T))^{1/4}$. These two layers unite and form the Stewartson layer (with $\delta_S=D\,Ek^{1/3}$) if stratification decreases \citep{barcilon}. For the present case $\delta_h > b-a$ holds, i.e. practically the whole measurement cavity lies within the ``hydrostatic'' domain. The buoyancy layer, however, is found to be only $\delta_b\approx 1$ mm thick, thus it is not resolved sufficiently by most of the models. 

The sensitivity of drift rates to the horizontal grid spacing was demonstrated with the INCA model. The phase speeds of baroclinic waves at two different rotation rates -- namely $\Omega=4$ rpm and $\Omega=9.5$ rpm -- were determined using two grids in both cases for comparison. The coarse and fine grids had minimum cell sizes of $\Delta x_{\rm min,(1)}=\Delta y_{\rm min,(1)}=1.5$ mm and $\Delta x_{\rm min,(2)}=\Delta y_{\rm min,(2)}=0.5$ mm, respectively. The obtained drift rates were: $c^{(1)}=0.097$ rad/s (coarse grid); $c^{(2)}=0.057$ rad/s (fine grid) at $\Omega=4$ rpm, and $c^{(1)}=0.025$ rad/s (coarse grid); $c^{(2)}=0.023$ rad/s (fine grid) at $\Omega=9.5$ rpm. Visibly, at the lower rotation rate (where the phase velocities of baroclinic waves are generally large) the refinement of the horizontal grid yielded slower wave propagation almost by a factor of two. In the case of the higher rotation rate this effect was manifestly smaller -- around 10\% -- in qualitative agreement with the drag-hypothesis: the drag itself is expected to be smaller too if the drift itself is slower. Thus, we can conclude that the grid resolution has marked effect on the simulated wave speeds, and to get a proper insight into the flow structure at the vicinity of the lateral sidewalls, one needs to apply grids that properly resolve the buoyancy layer.   

We also found marked connection between the spatial patterns of baroclinic waves and their drift rates, both experimentally and numerically. The aforementioned hysteresis that was observed in terms of the dominant wave number $m$ also manifested itself in the drift rates. In the cylFloit simulations, $m=3$ waves always propagated faster than their $m=4$ counterparts at a given rotation rate (within the hysteretic $\Omega$-regime). Similar behavior was noticed in the laboratory experiments too: a certain type of the $m=3$ waves was found to be faster than the $m=4$ waves of the same $\Omega$. However, in the laboratory, another type of three-fold symmetric ($m=3$) pattern appeared as well in the ``spin-up'' series, which was found to propagate at even smaller speed than the $m=4$ waves. Here the surface temperature pattern has undergone a ``topological'' reorganization: the meandering warm jet that separated the inner and outer domain in the ``fast'' $m=3$ waves has disconnected. This transition possibly opens the way for stronger radial temperature fluxes, therefore this new configuration may reduce the thermal wind (background flow) more effectively, thus yielding slower drift. Applying the same reasoning for the hysteresis of $m=4$ waves and the ``fast'' $m=3$ waves, it can be stated that among these, the $m=4$ mode exhibits larger radial heat flow.      
As far as the general heat flow is considered, \cite{rjh} showed that the Nusselt number $Nu$ in a baroclinic annulus exhibits a large drop at the transition from axisymmetric flow to the regime of regular waves, where -- compared to the abrupt change at the onset of baroclinic instability -- does not change markedly with the increasing $\Omega$. This plateau ends when the system reached the higher rotation rates where thy waves become irregular (this state was not studied in the present work), and is followed by a pronounced fall of $Nu$. Thus, it is to be noted, that the changes in heat flow that can be attributed to shape changes of regular beroclinic waves is rather small compared to the abrupt changes that arise outside the wave flow regime. We also remark, that -- as demonstrated in the experiments of \cite{fein} -- the drift rates are also highly sensitive to the upper boundary condition, that was not properly defined in the model equations.

The third main focus of our study was the statistical quantification of the structures of the surface temperature field and the analysis of their spatio-temporal variability. 
As a measure of the overall variability in the system, the time averaged temperature variance $\bar{\sigma}(\Omega)$ taken along the circular contour at mid radius $r_{\rm mid}$ was intended to serve as an order parameter that indicates the breaking of the axial symmetry (and, thus, the onset of baroclinic instability) with a marked jump at critical rotation rate $\Omega_{\rm crit}$. Indeed, in the numerical simulations $\bar{\sigma}\approx 0$ was detected in all cases where no dominant wave mode could be found (aside for the trivial $m=0$), implying the stability of the axially symmetric basic state. This was then followed by more than $10$ times larger variances at $\Omega > \Omega_{\rm crit}$. However, in the laboratory experiments the transition was not that apparent: even below $\Omega_{\rm crit}$ fluctuations appeared on the same order of magnitude as the $\bar{\sigma}$ values of higher rotation rates (though, smaller by a factor of $\approx 0.5$). This observation confirms our previous finding of spontaneous excitation of dispersive transient wave-like phenomena (coined ``weak waves'') that ``blur'' the boundary of instability in the parameter space \citep{our_npg}. This qualitative difference between numerics and experiments indicates the presence of non-modal transient growth of small temperature fluctuations in this sensitive regime \citep{torsten} unavoidable in the laboratory (see also the work of \cite{hoff} in the present volume). In the numerical results the temperature variance obtained at a few centimeters below the surface was found to be significantly larger (by a factor of $\approx 2$) than at the surface. This behavior, however, could not be verified experimentally with the applied measurement techniques.

In order to analyze smaller scale spatial structures, we calculated the Fourier spectra of the azimuthal temperature profiles along the circular contour at mid-radius $r_{\rm mid}$ for all time instants of a given experimental or numerical run, and their temporal average was considered as the characteristic spectral ``fingerprint'' of the investigated pattern. In the case of an $m$-fold symmetric baroclinic wave, besides the dominant wave number, its harmonics also appear in the spectra with finite amplitudes, as already demonstrated by \cite{james}. The amplitudes and the significance of the spectral peaks provide a measure of the importance of the regular smaller scale patterns. Typically, in the experimental data the amplitudes at the integer multiples of the dominant mode were markedly present, in many cases exhibiting comparable amplitudes to the dominant wave number corresponding to the overall rotational symmetry. In the cylFloit simulations however, the harmonics were not that pronounced. These smaller-scale patterns are attributed to the cold eddies outside and inside the meandering jet of the baroclinic wave. The fact that these structures could not be resolved accurately in the simulations may be due to one (or more) of the following reasons: (i) the cold eddies in the vicinity of the outer rim may be excited by shear instability involving the bouyancy layer, which was not resolved by most of the models, as discussed above; (ii) surface phenomena that are out of the scope of the studied governing equations may also be responsible, e.g. the ``wind'' stress that takes place at the free surface of the experimental tank as it rotates, or the presence of finite vertical heat fluxes at the top surface (note, that all the models included the $\nabla T \vec{e_z}|_{z=D}=0$ type no-flux boundary conditions, which certainly cannot be achieved in the experiment due to the free surface).

The azimuthal temperature variance patterns were also decomposed into sets of empirical orthogonal functions (EOFs). We found that in the experimental distribution of the ranked relative variances -- the normalized eigenvalues corresponding to the EOF modes -- typically follows a slowly decaying power-law type scaling, implying that a considerable part of the total variance is stored in the smaller scales (6-18 orthogonal modes were needed to cover 90\% of the total variance). In general, the numerically obtained distributions exhibited faster cut-offs towards the higher ranks, thus less small-scale variance. The practical absence of the correlated small-scale thermal fluctuations in the simulations supports the need for some subgrid-scale parametrization that takes into account the growth of temperature fluctuations that might play a significant role in the dynamics. These fluctuations can be caused by the aforementioned experimental impurities (or possibly induced by boundary layer effects) and ``inflated'' through the nonlinear interactions.

As a possible extension and continuation of this idea, the response of the system to small amplitude temporal and spatial thermal fluctuations (entering via the boundary conditions) could be analyzed numerically in a future research project. Such investigations -- if the above assumptions are correct -- can possibly lead to even more accurate numerical modeling and a deeper understanding of the dynamics in this set-up. Also, our future plans involve the extension of the presented benchmarking techniques to numerical methods that reach beyond the Boussinesq approximation (e.g. Low-Mach models) whose application may be wise in the larger $\Delta T$-regime.  

The results presented in this paper have clearly demonstrated that the relatively simple rotating annulus arrangement indeed provides a remarkable test bed to verify and tune numerical methods aiming to model large-scale atmospheric flows. The authors think that the presented pool of experimental and numerical data and the applied evaluation methods and ``test quantities'' will also prove useful benchmarks for similar studies in the future.       
     
\section*{Acknowledgements}

This work has been funded by the German Science Foundation (DFG)
and is part of the DFG priority program MetStr\"om (SPP 1276).
The experimental team (U.H., M.V., K.D.A. and C.E.) is grateful for the technical help of H.-J. Pflanz, R. St\"obel and Y. Wang throughout the period of the whole MetStr\"om program. 
The Hiflow$^3$ team (M.B., T.B. and V.H.) gratefully acknowledges 
the computing time granted by the John von Neumann Institute for Computing (NIC) 
and provided on the supercomputer JUROPA at J\"ulich Supercomputing Centre (JSC).
The work of the HiFlow$^3$ team was partially supported by grant HE4760/3-3 (DFG).
The cylFloit team (S.B. and U.A.) were also financed through DFG grant Ac71/4-2. 
The INCA team (S.R. and S.H.) was supported by DFG grant HI1273-1, and the computational resources were provided by the HLRS Stuttgart, under grant TIGRA.
The EULAG
numerical data were generated using resources of the Department of
Mathematics and Computer Science, Freie Universit\"at Berlin, Germany.
Computing time for the results obtained with LESOCC2 (C.H. and J.F.) was provided by ZIH at TU Dresden.
M.V. acknowledges the discussions with the members of the project group ``Physical Mechanisms of Global Environmental Processes'' (Budapest, Hungary, Grant number NK100296).


\bibliography{references_benchmark}

\begin{thebibliography}{}

\bibitem[{\sc Achatz} and {\sc Schmitz}(1997){{\sc Achatz} and {\sc
  Schmitz}}]{achatz}
{\sc Achatz, U.}, {\sc G.~Schmitz}, 1997:
\newblock On the closure problem in the reduction of complex atmospheric models
  by {PIP}s and {EOF}s: {A} comparison for the case of a two-layer model with
  zonally symmetric forcing.
\newblock -- J. Atmos. Sci. {\bf 54}, 2452--2474.

\bibitem[{\sc Barcilon} and {\sc Pedlosky}(1967){{\sc Barcilon} and {\sc
  Pedlosky}}]{barcilon}
{\sc Barcilon, V.}, {\sc J.~Pedlosky}, 1967:
\newblock A unified linear theory of homogeneous and stratified rotating
  fluids.
\newblock -- Journal of Fluid Mechanics {\bf 29}(03), 609--621.

\bibitem[{\sc Borchert} et~al.(2014){{\sc Borchert}, {\sc Achatz}, {\sc
  Remmler}, {\sc Hickel}, {\sc Harlander}, {\sc Vincze}, {\sc Alexandrov}, {\sc
  Rieper}, {\sc Heppelmann}, and {\sc Dolaptchiev}}]{borchert}
{\sc Borchert, S.}, {\sc U.~Achatz}, {\sc S.~Remmler}, {\sc S.~Hickel}, {\sc
  U.~Harlander}, {\sc M.~Vincze}, {\sc K.~D. Alexandrov}, {\sc F.~Rieper}, {\sc
  T.~Heppelmann}, {\sc S.~I. Dolaptchiev}, 2014:
\newblock Finite-volume models with implicit subgrid-scale parameterization for
  the differentially heated rotating annulus.
\newblock -- Meterol. Z., under review

\bibitem[{\sc Brezzi}(1974){{\sc Brezzi}}]{1974Brezzi}
{\sc Brezzi, F.}, 1974:
\newblock On the existence, uniqueness, and approximation of saddle point
  problems arising from {L}agrangian multipliers.
\newblock -- Revue fran\c{c}aise d'automatique, informatique, recherche
  op\'{e}rationnelle. Analyse Num\'{e}rique {\bf 8}(2), 129--151.

\bibitem[{\sc Chorin}(1968){{\sc Chorin}}]{chorin}
{\sc Chorin, A.~J.}, 1968:
\newblock Numerical solution of the {N}avier--{S}tokes equations.
\newblock -- Math. Comp. {\bf 22}, 745--762.

\bibitem[{\sc Cotter} et~al.(2002){{\sc Cotter}, {\sc Smolarkiewicz}, and {\sc
  Szczyrba}}]{Cotter2002}
{\sc Cotter, C.}, {\sc P.~Smolarkiewicz}, {\sc I.~Szczyrba}, 2002:
\newblock A viscoelastic fluid model for brain injuries.
\newblock -- International Journal for Numerical Methods in Fluids {\bf
  40}(1-2), 303--311.

\bibitem[{\sc Eady}(1949){{\sc Eady}}]{eady}
{\sc Eady, E.}, 1949:
\newblock Long waves and cyclone waves.
\newblock -- Tellus {\bf 1}(3), 33--52.

\bibitem[{\sc Elliott} and {\sc Smolarkiewicz}(2002){{\sc Elliott} and {\sc
  Smolarkiewicz}}]{Elliott2002}
{\sc Elliott, J.}, {\sc P.~Smolarkiewicz}, 2002:
\newblock Eddy resolving simulations of turbulent solar convection.
\newblock -- International Journal for Numerical Methods in Fluids {\bf 39}(9),
  855--864.

\bibitem[{\sc Fein}(1973){{\sc Fein}}]{fein}
{\sc Fein, J.~S.}, 1973:
\newblock An experimental study of the effects of the upper boundary condition
  on the thermal convection in a rotating, differentially heated cylindrical
  annulus of water.
\newblock -- Geophysical \& Astrophysical Fluid Dynamics {\bf 5}(1), 213--248.

\bibitem[{\sc Fr{\"o}hlich}(2006){{\sc Fr{\"o}hlich}}]{froehlich2006}
{\sc Fr{\"o}hlich, J.}, 2006:
\newblock Large Eddy Simulation turbulenter Str{\"o}mungen (in German)
\newblock -- Teubner Verlag, 414.

\bibitem[{\sc Fr{\"u}h} and {\sc Read}(1997){{\sc Fr{\"u}h} and {\sc
  Read}}]{read}
{\sc Fr{\"u}h, W.-G.}, {\sc P.~Read}, 1997:
\newblock Wave interactions and the transition to chaos of baroclinic waves in
  a thermally driven rotating annulus.
\newblock -- Philosophical Transactions of the Royal Society of London. Series
  A: Mathematical, Physical and Engineering Sciences {\bf 355}(1722), 101--153.

\bibitem[{\sc Fultz} et~al.(1959){{\sc Fultz}, {\sc Long}, {\sc Owens}, and
  {\sc Weil}}]{fultz}
{\sc Fultz, D.}, {\sc R.~R. Long}, {\sc G.~V. Owens}, {\sc J.~Weil}, 1959:
\newblock Studies of thermal convection in a rotating cylinder with some
  implications for large-scale atmospheric motions
\newblock -- American Meteorological Society.

\bibitem[{\sc Goldstein} et~al.(1993){{\sc Goldstein}, {\sc Handler}, and {\sc
  Sirovich}}]{Goldstein1993}
{\sc Goldstein, D.}, {\sc R.~Handler}, {\sc L.~Sirovich}, 1993:
\newblock Modeling a no-slip flow boundary with an external force field.
\newblock -- J. Comput. Phys. {\bf 105}(2), 354--366.

\bibitem[{\sc Grabowski} and {\sc Smolarkiewicz}(2002){{\sc Grabowski} and {\sc
  Smolarkiewicz}}]{Grabowski2002}
{\sc Grabowski, W.}, {\sc P.~Smolarkiewicz}, 2002:
\newblock A multiscale anelastic model for meteorological research.
\newblock -- Mon. Weather Rev. {\bf 130}, 939--956.

\bibitem[{\sc Grilli} et~al.(2012){{\sc Grilli}, {\sc Schmid}, {\sc Hickel},
  and {\sc Adams}}]{grilli}
{\sc Grilli, M.}, {\sc P.~J. Schmid}, {\sc S.~Hickel}, {\sc N.~A. Adams}, 2012:
\newblock Analysis of unsteady behaviour in shockwave turbulent boundary layer
  interaction.
\newblock -- J. Fluid Mech. {\bf 700}, 16--28.

\bibitem[{\sc Gy{\"u}re} et~al.(2007){{\sc Gy{\"u}re}, {\sc Bartos}, and {\sc
  J{\'a}nosi}}]{gyure}
{\sc Gy{\"u}re, B.}, {\sc I.~Bartos}, {\sc I.~M. J{\'a}nosi}, 2007:
\newblock Nonlinear statistics of daily temperature fluctuations reproduced in
  a laboratory experiment.
\newblock -- Physical Review E {\bf 76}(3), 037301.

\bibitem[{\sc Harlander} et~al.(2009){{\sc Harlander}, {\sc Faulwetter}, {\sc
  Alexandrov}, and {\sc Egbers}}]{uwe_ait}
{\sc Harlander, U.}, {\sc R.~Faulwetter}, {\sc K.~Alexandrov}, {\sc C.~Egbers},
  2009:
\newblock Estimating local instabilities for irregular flows in the
  differentially heated rotating annulus.
\newblock -- In: Advances in Turbulence XII, Springer, 163--166.

\bibitem[{\sc Harlander} et~al.(2011){{\sc Harlander}, {\sc von Larcher}, {\sc
  Wang}, and {\sc Egbers}}]{uwe_piv}
{\sc Harlander, U.}, {\sc von T.~Larcher}, {\sc Y.~Wang}, {\sc C.~Egbers},
  2011:
\newblock Piv-and ldv-measurements of baroclinic wave interactions in a
  thermally driven rotating annulus.
\newblock -- Experiments in fluids {\bf 51}(1), 37--49.

\bibitem[{\sc Harlander} et~al.(2012){{\sc Harlander}, {\sc Wenzel}, {\sc
  Alexandrov}, {\sc Wang}, and {\sc Egbers}}]{uwe_obst}
{\sc Harlander, U.}, {\sc J.~Wenzel}, {\sc K.~Alexandrov}, {\sc Y.~Wang}, {\sc
  C.~Egbers}, 2012:
\newblock Simultaneous piv and thermography measurements of partially blocked
  flow in a differentially heated rotating annulus.
\newblock -- Experiments in fluids {\bf 52}(4), 1077--1087.

\bibitem[{\sc Harlander} et~al.(2014){{\sc Harlander}, {\sc v.~Larcher}, {\sc
  Wright}, {\sc Hoff}, {\sc Alexandrov}, and {\sc Egbers}}]{uwe_agu}
{\sc Harlander, U.}, {\sc v.~T.~Larcher}, {\sc G.~B. Wright}, {\sc M.~Hoff},
  {\sc K.~D. Alexandrov}, {\sc C.~Egbers}, 2014:
\newblock Orthogonal decomposition methods to analyze piv, ldv and thermography
  data of a thermally driven rotating annulus laboratory experiment.
\newblock -- AGU GEOPHYSICAL MONOGRAPH SERIES, book title 'Modelling
  Atmospheric and Oceanic flows: insights from laboratory experiments and
  numerical simulations' {\bf accepted}.

\bibitem[{\sc Heuveline}(2010){{\sc Heuveline}}]{2010Heuveline_HiFlow3}
{\sc Heuveline, V.}, 2010:
\newblock {HiFlow3: A flexible and hardware-aware parallel finite element
  package}.
\newblock -- In: Proceedings of the 9th Workshop on Parallel/High-Performance
  Object-Oriented Scientific Computing, POOSC '10, 4:1--4:6, New York, NY, USA.
  ACM.

\bibitem[{\sc Heuveline} et~al.(2012){{\sc Heuveline}, {\sc Ketelaer}, {\sc
  Ronnas}, {\sc Schmidtobreick}, and {\sc Wlotzka}}]{emcl-preprint-2012-05}
{\sc Heuveline, V.}, {\sc E.~Ketelaer}, {\sc S.~Ronnas}, {\sc
  M.~Schmidtobreick}, {\sc M.~Wlotzka}, 2012:
\newblock {Scalability Study of HiFlow3 based on a Fluid Flow Channel
  Benchmark}.

\bibitem[{\sc Hickel} et~al.(2006){{\sc Hickel}, {\sc Adams}, and {\sc
  Domaradzki}}]{hickel_06}
{\sc Hickel, S.}, {\sc N.~A. Adams}, {\sc J.~A. Domaradzki}, 2006:
\newblock An adaptive local deconvolution method for implicit {LES}.
\newblock -- J. Comput. Phys. {\bf 213}, 413--436.

\bibitem[{\sc Hickel} et~al.(2008){{\sc Hickel}, {\sc Kempe}, and {\sc
  Adams}}]{hickel_08}
{\sc Hickel, S.}, {\sc T.~Kempe}, {\sc N.~A. Adams}, 2008:
\newblock Implicit large-eddy simulation applied to turbulent channel flow with
  periodic constrictions.
\newblock -- Theor. Comput. Fluid Dyn. {\bf 22}, 227--242.

\bibitem[{\sc Hignett} et~al.(1985){{\sc Hignett}, {\sc White}, {\sc Carter},
  {\sc Jackson}, and {\sc Small}}]{hignett}
{\sc Hignett, B.~P.}, {\sc A.~White}, {\sc R.~Carter}, {\sc W.~Jackson}, {\sc
  R.~Small}, 1985:
\newblock A comparison of laboratory measurements and numerical simulations of
  baroclinic wave flows in a rotating cylindrical annulus.
\newblock -- Quarterly Journal of the Royal Meteorological Society {\bf
  111}(467), 131--154.

\bibitem[{\sc Hinterberger} et~al.(2007){{\sc Hinterberger}, {\sc
  Fr{\"o}hlich}, and {\sc Rodi}}]{Hinterberger2007}
{\sc Hinterberger, C.}, {\sc J.~Fr{\"o}hlich}, {\sc W.~Rodi}, 2007:
\newblock Three-dimensional and depth-averaged large-eddy simulations of some
  shallow water flows.
\newblock -- Journal of Hydraulic Engineering {\bf 133}, 857--872.

\bibitem[{\sc Hoff} et~al.(2014){{\sc Hoff}, {\sc Harlander}, and {\sc
  Egbers}}]{hoff}
{\sc Hoff, M.}, {\sc U.~Harlander}, {\sc C.~Egbers}, 2014:
\newblock Empirical singular vectors of baroclinic flows deduced from
  experimental data of a differentially heated rotating annulus.
\newblock -- Meteorologische Zeitschrift {\bf submitted}.

\bibitem[{\sc James} et~al.(1981){{\sc James}, {\sc Jonas}, and {\sc
  Farnell}}]{james}
{\sc James, I.}, {\sc P.~Jonas}, {\sc L.~Farnell}, 1981:
\newblock A combined laboratory and numerical study of fully developed steady
  baroclinic waves in a cylindrical annulus.
\newblock -- Quarterly Journal of the Royal Meteorological Society {\bf
  107}(451), 51--78.

\bibitem[{\sc J{\'a}nosi} et~al.(2010){{\sc J{\'a}nosi}, {\sc Kiss}, {\sc
  Homonnai}, {\sc Pattanty{\'u}s-{\'A}brah{\'a}m}, {\sc Gy{\"u}re}, and {\sc
  T{\'e}l}}]{viki}
{\sc J{\'a}nosi, I.~M.}, {\sc P.~Kiss}, {\sc V.~Homonnai}, {\sc
  M.~Pattanty{\'u}s-{\'A}brah{\'a}m}, {\sc B.~Gy{\"u}re}, {\sc T.~T{\'e}l},
  2010:
\newblock Dynamics of passive tracers in the atmosphere: Laboratory experiments
  and numerical tests with reanalysis wind fields.
\newblock -- Physical Review E {\bf 82}(4), 046308.

\bibitem[{\sc Lorenz}(1963){{\sc Lorenz}}]{lorenz}
{\sc Lorenz, E.~N.}, 1963:
\newblock The mechanics of vacillation.
\newblock -- Journal of the Atmospheric Sciences {\bf 20}(5), 448--465.

\bibitem[{\sc Mason}(1975){{\sc Mason}}]{mason}
{\sc Mason, P.}, 1975:
\newblock Baroclinic waves in a container with sloping end walls.
\newblock -- Philosophical Transactions of the Royal Society of London. Series
  A, Mathematical and Physical Sciences {\bf 278}(1284), 397--445.

\bibitem[{\sc Mayer}(2007){{\sc Mayer}}]{2007ILUPP}
{\sc Mayer, J.}, 2007:
\newblock Ilu++: A new software package for solving sparse linear systems with
  iterative methods.
\newblock -- PAMM {\bf 7}(1), 2020123--2020124.

\bibitem[{\sc Meyer} et~al.(2010){{\sc Meyer}, {\sc Hickel}, and {\sc
  Adams}}]{meyer}
{\sc Meyer, M.}, {\sc S.~Hickel}, {\sc N.~Adams}, 2010:
\newblock Assessment of implicit large-eddy simulation with a conservative
  immersed interface method for turbulent cylinder flow.
\newblock -- International Journal of Heat and Fluid Flow {\bf 31}(3), 368 --
  377.

\bibitem[{\sc Miller} and {\sc Butler}(1991){{\sc Miller} and {\sc
  Butler}}]{miller_butler}
{\sc Miller, T.~L.}, {\sc K.~A. Butler}, 1991:
\newblock Hysteresis and the transition between axisymmetric flow and wave flow
  in the baroclinic annulus.
\newblock -- Journal of the atmospheric sciences {\bf 48}(6), 811--824.

\bibitem[{\sc Monahan} et~al.(2009){{\sc Monahan}, {\sc Fyfe}, {\sc Ambaum},
  {\sc Stephenson}, and {\sc North}}]{eof_review}
{\sc Monahan, A.~H.}, {\sc J.~C. Fyfe}, {\sc M.~H. Ambaum}, {\sc D.~B.
  Stephenson}, {\sc G.~R. North}, 2009:
\newblock Empirical orthogonal functions: The medium is the message..
\newblock -- Journal of Climate {\bf 22}(24).

\bibitem[{\sc Pfeffer} and {\sc Fowlis}(1968){{\sc Pfeffer} and {\sc
  Fowlis}}]{pfeffer}
{\sc Pfeffer, R.~L.}, {\sc W.~W. Fowlis}, 1968:
\newblock Wave dispersion in a rotating, differentially heated cylindrical
  annulus of fluid.
\newblock -- Journal of the Atmospheric Sciences {\bf 25}(3), 361--371.

\bibitem[{\sc Prusa} et~al.(2008){{\sc Prusa}, {\sc Smolarkiewicz}, and {\sc
  Wyszogrodzki}}]{Prusa2008}
{\sc Prusa, J.}, {\sc P.~Smolarkiewicz}, {\sc A.~Wyszogrodzki}, 2008:
\newblock Eulag, a computational model for multiscale flows.
\newblock -- Comput. Fluids. {\bf 37}, 1193--1207.

\bibitem[{\sc Randriamampianina} et~al.(2006){{\sc Randriamampianina}, {\sc
  Fr{\"u}h}, {\sc Read}, and {\sc Maubert}}]{anthony1}
{\sc Randriamampianina, A.}, {\sc W.-G. Fr{\"u}h}, {\sc P.~L. Read}, {\sc
  P.~Maubert}, 2006:
\newblock Direct numerical simulations of bifurcations in an air-filled
  rotating baroclinic annulus.
\newblock -- Journal of Fluid Mechanics {\bf 561}, 359--389.

\bibitem[{\sc Ravela} et~al.(2010){{\sc Ravela}, {\sc Marshall}, {\sc Hill},
  {\sc Wong}, and {\sc Stransky}}]{ravela}
{\sc Ravela, S.}, {\sc J.~Marshall}, {\sc C.~Hill}, {\sc A.~Wong}, {\sc
  S.~Stransky}, 2010:
\newblock A realtime observatory for laboratory simulation of planetary flows.
\newblock -- Experiments in fluids {\bf 48}(5), 915--925.

\bibitem[{\sc Rayer} et~al.(1998){{\sc Rayer}, {\sc Johnson}, and {\sc
  Hide}}]{rjh}
{\sc Rayer, Q.}, {\sc D.~Johnson}, {\sc R.~Hide}, 1998:
\newblock Thermal convection in a rotating fluid annulus blocked by a radial
  barrier.
\newblock -- Geophysical \& Astrophysical Fluid Dynamics {\bf 87}(3-4),
  215--252.

\bibitem[{\sc Read} et~al.(1997){{\sc Read}, {\sc Lewis}, and {\sc Hide}}]{rlh}
{\sc Read, P.}, {\sc S.~Lewis}, {\sc R.~Hide}, 1997:
\newblock Laboratory and numerical studies of baroclinic waves in an internally
  heated rotating fluid annulus: a case of wave/vortex duality?.
\newblock -- Journal of Fluid Mechanics {\bf 337}, 155--191.

\bibitem[{\sc Read}(2003){{\sc Read}}]{read_03}
{\sc Read, P.~L.}, 2003:
\newblock A combined laboratory and numerical study of heat transport by
  baroclinic eddies and axisymmetric flows.
\newblock -- Journal of Fluid Mechanics {\bf 489}, 301--323.

\bibitem[{\sc Read} et~al.(2008){{\sc Read}, {\sc Maubert}, {\sc
  Randriamampianina}, and {\sc Fr{\"u}h}}]{anthony2}
{\sc Read, P.~L.}, {\sc P.~Maubert}, {\sc A.~Randriamampianina}, {\sc W.-G.
  Fr{\"u}h}, 2008:
\newblock Direct numerical simulation of transitions towards structural
  vacillation in an air-filled, rotating, baroclinic annulus.
\newblock -- Physics of Fluids (1994-present) {\bf 20}(4), 044107.

\bibitem[{\sc Remmler} and {\sc Hickel}(2012){{\sc Remmler} and {\sc
  Hickel}}]{rh_12}
{\sc Remmler, S.}, {\sc S.~Hickel}, 2012:
\newblock Direct and large eddy simulation of stratified turbulence.
\newblock -- Int. J. Heat Fluid Flow {\bf 35}, 13--24.

\bibitem[{\sc Remmler} and {\sc Hickel}(2013){{\sc Remmler} and {\sc
  Hickel}}]{rh_13}
{\sc Remmler, S.}, {\sc S.~Hickel}, 2013:
\newblock Spectral structure of stratified turbulence: {D}irect numerical
  simulations and predictions by large eddy simulation.
\newblock -- Theor. Comput. Fluid Dyn. {\bf 27}, 319--336.

\bibitem[{\sc Rhie} and {\sc Chow}(1983){{\sc Rhie} and {\sc Chow}}]{Rhie1983}
{\sc Rhie, C.~M.}, {\sc W.~L. Chow}, 1983:
\newblock Numerical study of the turbulent flow past an airfoil with trailing
  edge separation.
\newblock -- AIAA Journal {\bf 21}, 1525--1532.

\bibitem[{\sc Schr{\"o}ttle} and {\sc D{\"o}rnbrack}(2013){{\sc Schr{\"o}ttle}
  and {\sc D{\"o}rnbrack}}]{Schroettle2013}
{\sc Schr{\"o}ttle, J.}, {\sc A.~D{\"o}rnbrack}, 2013:
\newblock Turbulence structure in a diabatically heated forest canopy composed
  of fractal pythagoras trees.
\newblock -- Theoretical and Computational Fluid Dynamics {\bf 27}(3-4),
  337--359.

\bibitem[{\sc Seelig} et~al.(2012){{\sc Seelig}, {\sc Harlander}, {\sc
  Faulwetter}, and {\sc Egbers}}]{torsten}
{\sc Seelig, T.}, {\sc U.~Harlander}, {\sc R.~Faulwetter}, {\sc C.~Egbers},
  2012:
\newblock Irregularity and singular vector growth in the differentially heated
  rotating annulus.
\newblock -- Theoret. Comp. Fluid Dynamics, published online, DOI {\bf 10}.

\bibitem[{\sc Shu}(1988){{\sc Shu}}]{shu}
{\sc Shu, C.-W.}, 1988:
\newblock Total-variation-diminishing time discretizations.
\newblock -- SIAM J. Sci. Stat. Comput. {\bf 9(6)}, 1073--1084.

\bibitem[{\sc Sitte} and {\sc Egbers}(2000){{\sc Sitte} and {\sc
  Egbers}}]{christoph_book}
{\sc Sitte, B.}, {\sc C.~Egbers}, 2000:
\newblock Higher order dynamics of baroclinic waves.
\newblock -- In: Physics of Rotating Fluids, Springer, 355--375.

\bibitem[{\sc Smolarkiewicz}(1991){{\sc Smolarkiewicz}}]{Smolarkiewicz1991}
{\sc Smolarkiewicz, P.}, 1991:
\newblock On forward-in-time differencing for fluids.
\newblock -- Monthly Weather Review {\bf 119}, 2505--2510.

\bibitem[{\sc Smolarkiewicz} and {\sc Margolin}(1997){{\sc Smolarkiewicz} and
  {\sc Margolin}}]{Smolarkiewicz1997}
{\sc Smolarkiewicz, P.}, {\sc L.~Margolin}, 1997:
\newblock On forward-in-time differencing for fluids: An
  eulerian/semi-lagrangian nonhydrostatic model for stratified flows.
\newblock -- Atmos-Ocean Special {\bf 35}, 127--157.

\bibitem[{\sc Smolarkiewicz} and {\sc Margolin}(1998){{\sc Smolarkiewicz} and
  {\sc Margolin}}]{Smolarkiewicz1998}
{\sc Smolarkiewicz, P.}, {\sc L.~Margolin}, 1998:
\newblock {MPDATA}: A positive definite solver for geophysical flows.
\newblock -- J. Comput. Phys. {\bf 140}, 459--480.

\bibitem[{\sc Smolarkiewicz} et~al.(2007){{\sc Smolarkiewicz}, {\sc Sharman},
  {\sc Weil}, {\sc Perry}, {\sc Heist}, and {\sc Bowker}}]{Smolarkiewicz2007}
{\sc Smolarkiewicz, P.}, {\sc R.~Sharman}, {\sc J.~Weil}, {\sc S.~Perry}, {\sc
  D.~Heist}, {\sc G.~Bowker}, 2007:
\newblock Building resolving large-eddy simulations and comparison with wind
  tunnel experiments.
\newblock -- Journal of Computational Physics {\bf 227}(1), 633 -- 653.

\bibitem[{\sc Vallis}(2006){{\sc Vallis}}]{realvallis}
{\sc Vallis, G.~K.}, 2006:
\newblock Atmospheric and oceanic fluid dynamics: fundamentals and large-scale
  circulation
\newblock -- Cambridge University Press.

\bibitem[{\sc Vettin}(1857){{\sc Vettin}}]{vettin}
{\sc Vettin, F.}, 1857:
\newblock \"uber den aufsteigen luftstr\"om, die entstehung des hagels und der
  wirbel-st\"urme.
\newblock -- Ann. Physik Chemie {\bf 102}, 246--255.

\bibitem[{\sc Vincze} et~al.(2014){{\sc Vincze}, {\sc Harlander}, {\sc von
  Larcher}, and {\sc Egbers}}]{our_npg}
{\sc Vincze, M.}, {\sc U.~Harlander}, {\sc von T.~Larcher}, {\sc C.~Egbers},
  2014:
\newblock An experimental study of regime transitions in a differentially
  heated baroclinic annulus with flat and sloping bottom topographies.
\newblock -- Nonlinear Processes in Geophysics {\bf 21}(1), 237--250.

\bibitem[{\sc von Larcher} and {\sc D\"ornbrack}(2014){{\sc von Larcher} and
  {\sc D\"ornbrack}}]{thomas_mz}
{\sc von Larcher, T.}, {\sc A.~D\"ornbrack}, 2014:
\newblock Eulag model simulations of baroclinic driven flows in a thermally
  driven rotating annulus, meteorologische zeitschrift.
\newblock -- Meteorologische Zeitschrift {\bf submitted}.

\bibitem[{\sc Von~Larcher} et~al.(2005){{\sc Von~Larcher}, {\sc Egbers}, and
  {\sc others}}]{thomas_npg}
{\sc Von~Larcher, T.}, {\sc C.~Egbers}, {\sc others}, 2005:
\newblock Experiments on transitions of baroclinic waves in a differentially
  heated rotating annulus.
\newblock -- Nonlinear Processes in Geophysics {\bf 12}(6), 1033--1041.

\bibitem[{\sc von Larcher} et~al.(2013){{\sc von Larcher}, {\sc Fournier}, and
  {\sc Hollerbach}}]{thomas_slope}
{\sc von Larcher, T.}, {\sc A.~Fournier}, {\sc R.~Hollerbach}, 2013:
\newblock The influence of a sloping bottom endwall on the linear stability in
  the thermally driven baroclinic annulus with a free surface.
\newblock -- Theoretical and Computational Fluid Dynamics {\bf 27}(3-4),
  433--451.

\bibitem[{\sc von Storch} and {\sc Navarra}(1999){{\sc von Storch} and {\sc
  Navarra}}]{eof_book}
{\sc von Storch, H.}, {\sc A.~Navarra}, 1999:
\newblock Analysis of climate variability: applications of statistical
  techniques
\newblock -- Springer.

\bibitem[{\sc Williams}(1971){{\sc Williams}}]{gp_williams}
{\sc Williams, G.~P.}, 1971:
\newblock Baroclinic annulus waves.
\newblock -- Journal of Fluid Mechanics {\bf 49}(03), 417--449.

\bibitem[{\sc Williams} et~al.(2010){{\sc Williams}, {\sc Read}, and {\sc
  Haine}}]{williams}
{\sc Williams, P.~D.}, {\sc P.~L. Read}, {\sc T.~W. Haine}, 2010:
\newblock Testing the limits of quasi-geostrophic theory: application to
  observed laboratory flows outside the quasi-geostrophic regime.
\newblock -- Journal of Fluid Mechanics {\bf 649}, 187--203.

\bibitem[{\sc Williamson}(1980){{\sc Williamson}}]{williamson_1980}
{\sc Williamson, J.}, 1980:
\newblock {L}ow-storage {R}unge-{K}utta schemes.
\newblock -- J. Comput. Phys. {\bf 35}, 48--56.

\bibitem[{\sc Young} and {\sc Read}(2008){{\sc Young} and {\sc
  Read}}]{young_read_08}
{\sc Young, R.}, {\sc P.~Read}, 2008:
\newblock Breeding and predictability in the baroclinic rotating annulus using
  a perfect model.
\newblock -- Nonlinear Processes in Geophysics {\bf 15}(3).

\bibitem[{\sc Young} and {\sc Read}(2013){{\sc Young} and {\sc
  Read}}]{young_read_12}
{\sc Young, R.}, {\sc P.~Read}, 2013:
\newblock Data assimilation in the laboratory using a rotating annulus
  experiment.
\newblock -- Quarterly Journal of the Royal Meteorological Society {\bf
  139}(675), 1488--1504.

\bibitem[{\sc Zhu} and {\sc Rodi}(1992){{\sc Zhu} and {\sc Rodi}}]{Zhu1992b}
{\sc Zhu, J.}, {\sc W.~Rodi}, 1992:
\newblock Computation of axisymmetric confined jets in a diffuser.
\newblock -- International Journal for Numerical Methods in Fluids {\bf 14},
  241--251.

\end{thebibliography}
\makebibliography

\clearpage




\end{document}